\documentclass[review]{elsarticle}

\usepackage{mathtools}
\usepackage{hyperref}
\usepackage{booktabs} 
\usepackage{multirow}
\usepackage{xcolor}
\usepackage{diagbox}

\journal{Journal of Information Processing and Management}

\begin{document}

\begin{frontmatter}

\title{A Deep Look into Neural Ranking Models for Information Retrieval}

\author[UCAS,ICT]{Jiafeng Guo}
\author[UCAS,ICT]{Yixing Fan}
\author[UCAS,ICT]{Liang Pang}
\author[UMASS]{Liu Yang}
\author[UMASS]{Qingyao Ai}
\author[UMASS]{Hamed Zamani}
\author[UCAS,ICT]{Chen Wu}
\author[UMASS]{W. Bruce Croft}
\author[UCAS,ICT]{Xueqi Cheng}
\address[UCAS]{University of Chinese Academy of Sciences, Beijing, China}
\address[ICT]{CAS Key Lab of Network Data Science and Technology, Institute of Computing Technology, Chinese Academy of Sciences, Beijing, China}
\address[UMASS]{Center for Intelligent Information Retrieval, University of Massachusetts Amherst, Amherst, MA, USA}




\begin{abstract}
Ranking models lie at the heart of research on information retrieval (IR). During the past decades, different techniques have been proposed for constructing ranking models, from traditional heuristic methods, probabilistic methods, to modern machine learning methods. Recently, with the advance of deep learning technology, we have witnessed a growing body of work in applying shallow or deep neural networks to the ranking problem in IR, referred to as neural ranking models in this paper. The power of neural ranking models lies in the ability to learn from the raw text inputs for the ranking problem to avoid many limitations of hand-crafted features. Neural networks have sufficient capacity to model complicated tasks, which is needed to handle the complexity of relevance estimation in ranking. Since there have been a large variety of neural ranking models proposed, we believe it is the right time to summarize the current status,  learn from existing methodologies, and gain some insights for  future development. In contrast to existing reviews, in this survey, we will take a deep look into the neural ranking models from different dimensions to analyze their underlying assumptions, major design principles, and learning strategies. We compare these models through benchmark tasks to obtain a comprehensive empirical understanding of the existing techniques. We will also discuss what is missing in the current literature and what are the promising and desired future directions.
\end{abstract}

\begin{keyword}
neural ranking model \sep information retrieval \sep survey
\MSC[2010] 00-01\sep  99-00
\end{keyword}

\end{frontmatter}


\section{Introduction}
Information retrieval is a core task in many real-world applications, such as digital libraries, expert finding, Web search, and so on. Essentially, IR is the activity of obtaining some information resources relevant to an information need from within large collections. As there might be a variety of relevant resources, the returned results are typically ranked with respect to some relevance notion. This ranking of results is a key difference of IR from other problems. Therefore, research on ranking models has always been at the heart of IR.

Many different ranking models have been proposed over the past decades, including vector space models \cite{Salton1975}, probabilistic models \cite{robertson1976relevance}, and learning to rank (LTR) models \cite{Liu2009LTR,Li2011LTR}. Existing techniques, especially the LTR models, have already achieved great success in many IR applications, e.g., modern Web search engines like Google\footnote{http://google.com} or Bing\footnote{http://bing.com}. There is still, however, much room for improvement in the effectiveness of these techniques for more complex retrieval tasks.

In recent years, deep neural networks have led to exciting breakthroughs in speech recognition \cite{Hinton2012speech}, computer vision \cite{Krizhevsky2012ImageNet, LeCun2015Nature}, and natural language processing (NLP) \cite{goldberg2017neural,bahdanau2014neural}. These models have been shown to be effective at learning abstract representations from the raw input, and have sufficient model capacity to tackle difficult learning problems. Both of these are desirable properties for ranking models in IR. On one hand, most existing LTR models rely on hand-crafted features, which are usually time-consuming to design and often over-specific in definition. It would be of great value if ranking models could learn the useful ranking features automatically. On the other hand, relevance, as a key notion in IR, is often vague in definition and difficult to estimate since relevance judgments are based on a complicated human cognitive process. Neural models with sufficient model capacity have more potential for learning such complicated tasks than traditional shallow models. Due to these potential benefits and along with the expectation that similar successes with deep learning could be achieved in IR \cite{Craswell2016neuralIR}, we have witnessed substantial growth of work in applying neural networks for constructing ranking models in both academia and industry in recent years. Note that in this survey, we focus on neural ranking models for textual retrieval, which is central to IR, but not the only mode that neural models can be used for \cite{Wan2014deep,brenner2018end}.


Perhaps the first successful model of this type is the Deep Structured Semantic Model (DSSM) \cite{Huang2013} introduced in 2013, which is a neural ranking model that directly tackles the ad-hoc retrieval task. In the same year, Lu and Li \cite{Lu2013} proposed DeepMatch, which is a deep matching method applied to the Community-based Question Answering (CQA) and micro-blog matching tasks. Note that at the same time or even before this work, there were a number of studies focused on learning low-dimensional representations of texts with neural models \cite{salakhutdinov2009semantic, Mikolov2013distributed} and using them either within traditional IR models or with some new similarity metrics for ranking tasks. However, we would like to refer to those methods as representation learning models rather than neural ranking models, since they did not directly construct the ranking function with neural networks. Later, between 2014 and 2015, work on neural ranking models began to grow, such as new variants of DSSM \cite{Huang2013}, ARC I and ARC II \cite{Hu2014}, MatchPyramid \cite{Pang2016}, and so on. Most of this research focused on short text ranking tasks, such as TREC QA tracks and Microblog tracks \cite{Severyn2015}. Since 2016, the study of neural ranking models has bloomed, with significant work volume, deeper and more rigorous discussions, and much wider applications \cite{Onal2018neural}. For example, researchers began to discuss the practical effectiveness of neural ranking models on different ranking tasks \cite{Guo2016DRMM, cohen2016adaptability}. Neural ranking models have been applied to ad-hoc retrieval \cite{Mitra2017DUET, Hui2017}, community-based QA \cite{Qiu2015}, conversational search \cite{Yan2016learningtoresponse}, and so on. Researchers began to go beyond the architecture of neural ranking models, paying attention to new training paradigms of neural ranking models \cite{Dehghani:2017}, alternate indexing schemes for neural representations \cite{Zamani:2018:SNRM}, integration of external knowledge \cite{Xiong2017Word,Yang2018}, and other novel uses of neural approaches for IR tasks \cite{fan2017learning,tang2018deeptilebars}.

Up to now, we have seen exciting progress on neural ranking models. In academia, several neural ranking models learned from scratch can already outperform state-of-the-art LTR models with tens of hand-crafted features \cite{Pang2017, Fan2018}. Workshops and tutorials on this topic have attracted extensive interest in the IR community \cite{Craswell2016neuralIR, Craswell2017neuralIR}. Standard benchmark datasets \cite{Nguyen2016, wikiqa2015}, evaluation tasks \cite{trec2017car}, and open-source toolkits \cite{fan2017matchzoo} have been created to facilitate research and rigorous comparison. Meanwhile, in industry, we have also seen models such as DSSM put into a wide range of practical usage in the enterprise \cite{Gao2014tutorial}. Neural ranking models already generate the most important features for modern search engines. However, beyond these exciting results, there is still a long way to go for neural ranking models: 1) Neural ranking models have not had the level of breakthroughs achieved by neural methods in speech recognition or computer vision; 2) There is little understanding and few guidelines on the design principles of neural ranking models; 3) We have not identified the special capabilities of neural ranking models that go beyond traditional IR models. Therefore, it is the right moment to take a look back, summarize the current status, and gain some insights for future development.

There have been some related surveys on neural approaches to IR (neural IR for short). For example, Onal et al.\cite{Onal2018neural} reviewed the current landscape of neural IR research, paying attention to the application of neural methods to different IR tasks. Mitra and Craswell \cite{mitra2017neural} gave an introduction to neural information retrieval. In their booklet, they talked about fundamentals of text retrieval, and briefly reviewed IR methods employing pre-trained embeddings and neural networks. In contrast to this work, this survey does not try to cover every aspect of neural IR, but will focus on and take a deep look into ranking models with deep neural networks. Specifically, we formulate the existing neural ranking models under a unified framework, and review them from different dimensions to understand their underlying assumptions, major design principles, and learning strategies. We also compare representative neural ranking models through benchmark tasks to obtain a comprehensive empirical understanding. We hope these discussions will help researchers in neural IR learn from previous successes and failures, so that they can develop better neural ranking models in the future. In addition to the model discussion, we also introduce some trending topics in neural IR, including indexing schema, knowledge integration, visualized learning, contextual learning and model explanation. Some of these topics are important but have not been well addressed in this field, while others are very promising directions for future research.  


In the following, we will first introduce some typical textual IR tasks addressed by neural ranking models in Section 2. We then provide a unified formulation of neural ranking models in Section 3. From section 4 to 6, we review the existing models with regard to different dimensions as well as making empirical comparisons between them. We discuss trending topics in Section 7 and conclude the paper in Section 8.



\section{Major Applications of Neural Ranking Models}
\label{sec:application}

In this section, we describe several major textual IR applications where neural ranking models have been adopted and studied in the literature, including ad-hoc retrieval, question answering, community question answering, and automatic conversation. There are other applications where neural ranking models have been or could be applied, e.g., product search \cite{brenner2018end}, sponsored search \cite{Grbovic2015sponsored}, and so on. However, due to page limitations, we will not include these tasks in this survey.


\subsection{Ad-hoc Retrieval}
\label{sec:task_ad_hoc}
Ad-hoc retrieval is a classic retrieval task in which the user specifies his/her information need through a query which initiates a search (executed by the information system) for documents that  are likely to be relevant to the user. The term \textit{ad-hoc} refers to the scenario where documents in the collection remain relatively static while new queries are submitted to the system continually~\cite{Baeza2011}. The retrieved documents are typically returned as a ranking list through a ranking model where those at the top of the ranking are more likely to be relevant.

There has been a long research history on ad-hoc retrieval, with several well recognized characteristics and challenges associated with the task. A major characteristic of ad-hoc retrieval is the heterogeneity of the query and the documents. The query comes from a search user with potentially unclear intent and is usually very short, ranging from a few words to a few sentences \cite{mitra2017neural}. The documents are typically from a different set of authors and have longer text length, ranging from multiple sentences to many paragraphs. Such heterogeneity leads to the critical vocabulary mismatch problem \cite{Furnas1987mismatch,Zhao2010term}. Semantic matching, meaning matching words and phrases with similar meanings, could alleviate the problem, but exact matching is indispensable especially with rare terms \cite{Guo2016DRMM}. Such heterogeneity also leads to diverse relevance patterns. Different hypotheses, e.g. verbosity hypothesis and scope hypothesis \cite{Robertson:1994:SEA:188490.188561},  have been proposed considering the matching of a short query against a long document. The \textit{relevance} notion in ad-hoc retrieval is inherently vague in definition and highly user dependent, making relevance assessment a very challenging problem.


For the evaluation of different neural ranking models on the ad-hoc retrieval task, a large variety of TREC collections have been used. Specifically, retrieval experiments have been conducted over neural ranking models based on TREC collections such as Robust~\cite{Guo2016DRMM, Pang2016}, ClueWeb~\cite{Guo2016DRMM}, GOV2~\cite{Pang2017, Fan2018} and Microblog \cite{Pang2017}, as well as logs such as the AOL log \cite{Dehghani:2017} and the Bing Search log \cite{Huang2013, Shen2014, Palangi2016, Mitra2017DUET}. Recently, a new large scale dataset has been released, called the NTCIR WWW Task~\cite{zheng2018sogou}, which is suitable for experiments on neural ranking models.



\subsection{Question Answering}
\label{sec:task_qa}

Question-answering (QA) attempts to automatically answer questions posed by users in natural languages based on some information resources. 
The questions could be from a closed or open domain \cite{molla2007question}, while the information resources could vary from structured data (e.g., knowledge base) to unstructured data (e.g., documents or Web pages) \cite{Moschitti2016}. 
There have been a variety of task formats for QA, including multiple-choice selection \cite{mctest2013}, answer passage/sentence retrieval \cite{Voorhees2000,wikiqa2015},  answer span locating \cite{Rajpurkar2016}, and answer synthesizing from multiple sources \cite{mitra2016}. However, some of the task formats are usually not treated as an IR problem. For example, multiple-choice selection is typically formulated as a classification problem while answer span locating is usually studied under the machine reading comprehension topic.
In this survey, therefore, we focus on answer passage/sentence retrieval as it can be formulated as a typical IR problem and addressed by neural ranking models. Hereafter, we will refer to this specific task as QA for simplicity.


Compared with ad-hoc retrieval, QA shows reduced heterogeneity between the question and the answer passage/sentence. On one hand, the question is usually in natural language, which is longer than keyword queries and clearer in intent description. On the other hand, the answer passages/sentences are usually much shorter text spans than documents (e.g., the answer passage length of WikiPassageQA data is about 133 words \cite{Cohen2018WikiPassageQA}), leading to more concentrated topics/semantics. However, vocabulary mismatch is still a basic problem in QA. The notion of relevance is relatively clear in QA, i.e., whether the target passage/sentence answers the question, but assessment is challenging. Ranking models need to capture the patterns expected in the answer passage/sentence based on the intent of the question, such as the matching of the context words, the existence of the expected answer type, and so on.  



For the evaluation of QA tasks, several benchmark data sets have been developed, including TREC QA~\cite{Voorhees2000}, WikiQA~\cite{wikiqa2015}, WebAP~\cite{Keikha2014,Yang2016}, InsuranceQA~\cite{Feng15InsuranceQA}, WikiPassageQA~\cite{Cohen2018WikiPassageQA} and MS MARCO~\cite{Nguyen2016}. A variety of neural ranking models \cite{Wang2015BLSTM, Severyn2015, Yang2016aNMM, Qiu2015, Lu2013} have been tested on these data sets.


\subsection{Community Question Answering}
\label{sec:task_cqa}
Community question answering (CQA) aims to find answers to users' questions based on existing QA resources in CQA websites, such as Quora~\footnote{https://www.quora.com/}, Yahoo! Answers~\footnote{https://answers.yahoo.com}, Stack Overflow~\footnote{https://www.stackoverflow.com}, and Zhihu~\footnote{https://zhihu.com}. As a retrieval task, CQA can be further divided into two categories. The first is to directly retrieval answers from the answer pool, which is similar to the above QA task with some additional user behavioral data (e.g., upvotes/downvotes)~\cite{Yang2013}. So we will not discuss this format here again. The second is to retrieve similar questions from the question pool, based on the assumption that answers to similar question could answer new questions. Unless otherwise noted, we will refer to the second task format as CQA.


Since it involves the retrieval of similar questions, CQA is significantly different from the previous two tasks due to the homogeneity between the input question and target question. Specifically, both input and target questions are short natural language sentences (e.g. the question length in Yahoo! Answers is between 9 and 10 words on average \cite{Shtok2012}), describing users' information needs. Relevance in CQA refers to semantic equivalence/similarity, which is clear and symmetric in the sense that the two questions are exchangeable in the relevance definition. However, vocabulary mismatch is still a challenging problem as both questions are short and there exist different expressions for the same intent.


For evaluation of the CQA task, a large variety of data sets have been released for research. The well-known data sets include the Quora Dataset\footnote{https://data.quora.com/First-Quora-Dataset-Release-Question-Pairs}, Yahoo! Answers Dataset~\cite{Qiu2015} and  SemEval-2017 Task3~\cite{SemEval2017}. The recent proposed datasets include CQADupStack\footnote{https://github.com/D1Doris/CQADupStack}~\cite{hoogeveen2015cqadupstack}, ComQA\footnote{http://qa.mpi-inf.mpg.de/comqa}~\cite{abujabal2018comqa} and LinkSO~\cite{liu2018linkso}. 
A variety of neural ranking models \cite{Wang2017BiMPM, Pang2016, Wan2016MatchSRNN, Chen2018RIMatch, Qiu2015} have been tested on these data sets.

\subsection{Automatic Conversation}
\label{sec:task_automatic_conversation}

Automatic conversation (AC) aims to create an automatic human-computer dialog process for the purpose of question answering, task completion, and social chat (i.e., chit-chat) \cite{Gao2018}. In general, AC could be formulated either as an IR problem that aims to rank/select a proper response from a dialog repository \cite{Ji2014} or a generation problem that aims to generate an appropriate response with respect to the input utterance \cite{Ritter2011}. In this paper, we restrict AC to the social chat task with the IR formulation, since question answering has already been covered in the above QA task and task completion is usually not taken as an IR problem. From the perspective of conversation context, the IR-based AC could be further divided into single-turn conversation\cite{Wang2013} or multi-turn conversation~\cite{Wu2017SMN}.


When focusing on social chat, AC also shows homogeneity similar to CQA. That is, both the input utterance and the response are short natural language sentences (e.g., the utterance length of Ubuntu Dialog Corpus is between 10 to 11 words on average and the median conversation length of it is 6 words \cite{Lowe2015}). Relevance in AC refers to certain semantic correspondence (or coherent structure) which is broad in definition, e.g., given an input utterance ``OMG I got myopia at such an `old' age'', the response could range from general (e.g., ``Really?'') to specific (e.g., ``Yeah. Wish a pair of glasses as a gift'')~\cite{Yan2016learningtoresponse}. Therefore, vocabulary mismatch is no longer the central challenge in AC, as we can see from the example that a good response does not require semantic matching between the words. Instead, it is critical to model correspondence/coherence and avoid general trivial responses.


For the evaluation of different neural ranking models on the AC task, several conversation collections have been collected from social media such as forums, Twitter and Weibo. Specifically, experiments have been conducted over neural ranking models based on collections such as Ubuntu Dialog Corpus (UDC)~\cite{Wu2017SMN, Zhou2016multiview, Yang2017neural}, Sina Weibo dataset~\cite{Wang2013, Yan2016learningtoresponse, Yan2016shall, Yan2017joint}, MSDialog~\cite{Qu2018,Yang2018,Qu2019} and the "campaign" NTCIR STC~\cite{Shang2016}.



\section{A Unified Model Formulation}


Neural ranking models are mostly studied within the LTR framework. In this section, we give a unified formulation of neural ranking models from a generalized view of LTR problems.


Suppose that $\mathcal{S}$ is the \textit{generalized} query set, which could be the set of search queries, natural language questions or input utterances, and $\mathcal{T}$ is the \textit{generalized} document set, which could be the set of documents, answers or responses. Suppose that $\mathcal{Y}=\{1,2,\cdots, l\}$ is the label set where labels represent grades. There exists a total order between the grades $l\succ l-1 \succ \cdots \succ 1$, where $\succ$ denotes the order relation. Let $s_i \in \mathcal{S}$ be the $i$-th query, $T_i = \{t_{i,1},t_{i,2},\cdots,t_{i,n_i}\}\in \mathcal{T}$ be the set of documents associated with the query $s_i$, and $\mathbf{y}_i=\{y_{i,1},y_{i,2},\cdots,y_{i,n_i}\}$ be the set of labels associated with query $s_i$, where $n_i$ denotes the size of $T_i$ and $\mathbf{y}_i$ and $y_{i,j}$ denotes the relevance degree of $t_{i,j}$ with respect to $s_i$. 
Let $\mathcal{F}$ be the function class and $f(s_i, t_{i,j}) \in \mathcal{F}$ be a ranking function which associates a relevance score with a query-document pair.
Let $L(f;s_i,t_{i,j},\mathbf{y}_{i,j})$ be the loss function defined on prediction of $f$ over the query-document pair and their corresponding label. So a generalized LTR problem is to find the optimal ranking function $f^*$ by minimizing the loss function over some labeled dataset
\begin{equation}
    f^*= \arg \min \sum_{i}\sum_{j} L(f;s_i,t_{i,j},y_{i,j})
\end{equation}




Without loss of generality, the ranking function $f$ could be further abstracted by the following unified formulation
\begin{equation}\label{Eq.unified_formulation}
	f(s, t) = g(\psi(s), \phi(t), \eta(s,t))
\end{equation}
where $s$ and $t$ are two input texts, $\psi$, $\phi$ are representation functions which extract features from $s$ and $t$ respectively, $\eta$ is the interaction function which extracts features from $(s,t)$ pair, and $g$ is the evaluation function which computes the relevance score based on the feature representations.

Note that for traditional LTR approaches~\cite{Liu2009LTR}, functions $\psi$, $\phi$ and $\eta$ are usually set to be fixed functions (i.e., manually defined feature functions). The evaluation function $g$ can be any machine learning model, such as logistic regression or gradient boosting decision tree , which could be learned from the training data. For neural ranking models, in most cases, all the functions $\psi$, $\phi$, $\eta$ and $g$ are encoded in the network structures so that all of them can be learned from training data. 

In traditional LTR approaches, the inputs $s$ and $t$ are usually raw texts. In neural ranking models, we consider that the inputs could be either raw texts or word embeddings. In other words, embedding mapping is considered as a basic input layer, not included in $\psi$, $\phi$ and $\eta$. 

\section{Model Architecture}

Based on the above unified formulation, here we review existing neural ranking model architectures to better understand their basic assumptions and design principles.

\subsection{Symmetric vs. Asymmetric Architectures}
\label{sec:symm_nonsymm}

Starting from different underlying assumptions over the input texts $s$ and $t$, two major architectures emerge in neural ranking models, namely symmetric architecture and asymmetric architecture.

\textbf{Symmetric Architecture}: The inputs $s$ and $t$ are assumed to be homogeneous, so that symmetric network structure could be applied over the inputs.  
Note here symmetric structure means that the inputs $s$ and $t$ can exchange their positions in the input layer without affecting the final output. Specifically, there are two representative symmetric structures, namely siamese networks and symmetric interaction networks.

\textit{Siamese networks} literally imply symmetric structure in the network architecture. Representative models include DSSM~\cite{Huang2013}, CLSM~\cite{Shen2014} and LSTM-RNN~\cite{Palangi2016}. For example, DSSM represents two input texts with a unified process including the letter-trigram  mapping followed by the multi-layer perceptron (MLP) transformation, i.e., function $\phi$ is the same as function $\psi$. After that a cosine similarity function is applied to evaluate the similarity between the two representations, i.e., function $g$ is symmetric. Similarly, CLSM~\cite{Shen2014} replaces the representation functions $\psi$ and $\phi$ by two identical convolutional neural networks (CNNs) in order to capture the local word order information. LSTM-RNN~\cite{Palangi2016} replaces $\psi$ and $\phi$ by two identical long short-term memory (LSTM) networks in order to capture the long-term dependence between words.

\textit{Symmetric interaction networks}, as shown by the name, employ a symmetric interaction function to represent the inputs. Representative models include DeepMatch~\cite{Lu2013}, Arc-II~\cite{Hu2014}, MatchPyramid~\cite{Pang2016} and Match-SRNN~\cite{Wan2016MatchSRNN}. For example, Arc-II defines an interaction function $\eta$ over $s$ and $t$ by computing similarity (i.e., weighted sum) between every n-gram pair from $s$ and $t$, which is symmetric in nature. After that, several convolutional and max-pooling layers are leveraged to obtain the final relevance score, which is also symmetric over $s$ and $t$. MatchPyramid defines a symmetric interaction function $\eta$ between every word pair from $s$ and $t$ to capture fine-grained interaction signals. It then leverages a symmetric evaluation function $g$, i.e., several 2D CNNs and a dynamic pooling layer, to produce the relevance score.
A similar process can be found in DeepMatch and Match-SRNN.

Symmetric architectures, with the underlying homogeneous assumption, can fit well with the CQA and AC tasks, where $s$ and $t$ usually have similar lengths and similar forms (i.e., both are natural language sentences). They may sometimes work for the ad-hoc retrieval or QA tasks if one only uses document titles/snippets \cite{Huang2013} or short answer sentences \cite{Yang2016aNMM} to reduce the heterogeneity between the two inputs.

\textbf{Asymmetric Architecture:} The inputs $s$ and $t$ are assumed to be heterogeneous, so that asymmetric network structures should be applied over the inputs. Note here asymmetric structure means if we change the position of the inputs $s$ and $t$ in the input layer, we will obtain totally different output. 
Asymmetric architectures have been introduced mainly in the ad-hoc retrieval task \cite{Huang2013, Pang2017}, due to the inherent heterogeneity between the query and the document as discussed in Section \ref{sec:task_ad_hoc}. Such structures may also work for the QA task where answer passages are ranked against natural language questions \cite{Dai2018Conv-KNRM}. 

Here we take the ad-hoc retrieval scenario as an example to analyze the asymmetric architecture. We find there are three major strategies used in the asymmetric architecture to handle the heterogeneity between the query and the document, namely query split, document split, and joint split.

\begin{figure}
    \begin{minipage}[t]{0.3\linewidth}
        \centerline{
        \includegraphics[width=0.8\linewidth]{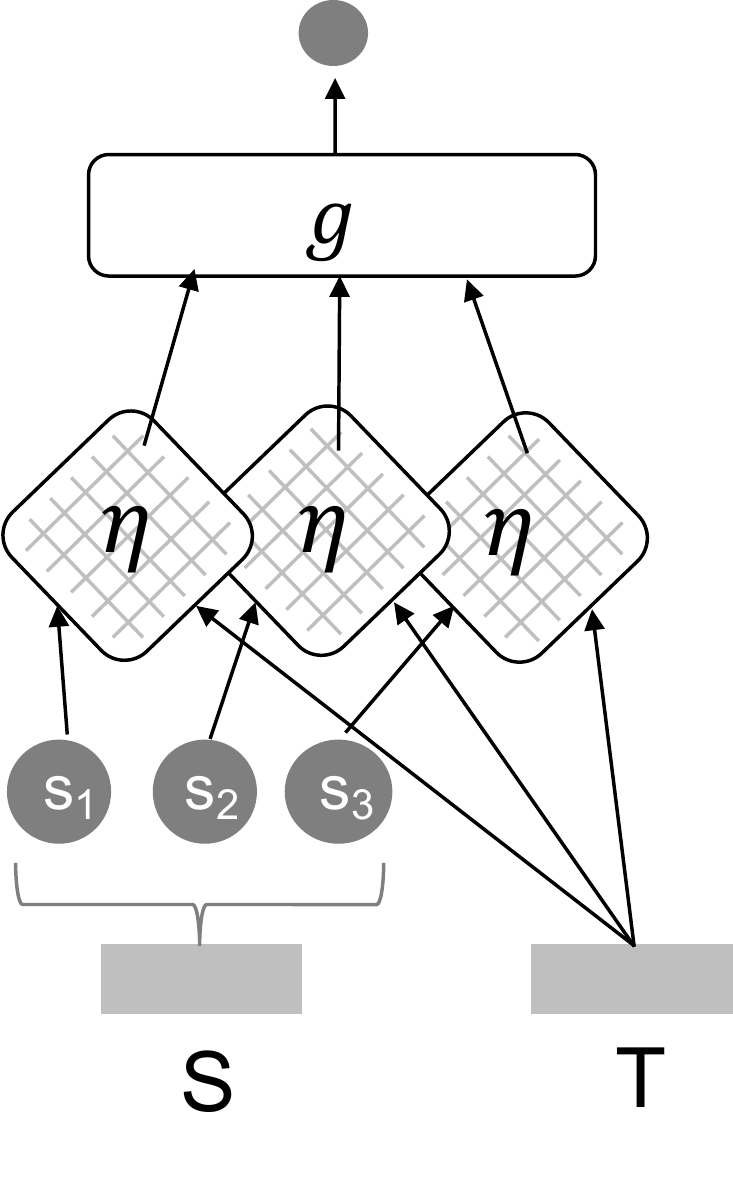}}
        \centerline{\footnotesize{(a) Query Split}}
    \end{minipage}
    \begin{minipage}[t]{0.3\linewidth}
        \centerline{
        \includegraphics[width=0.97\linewidth]{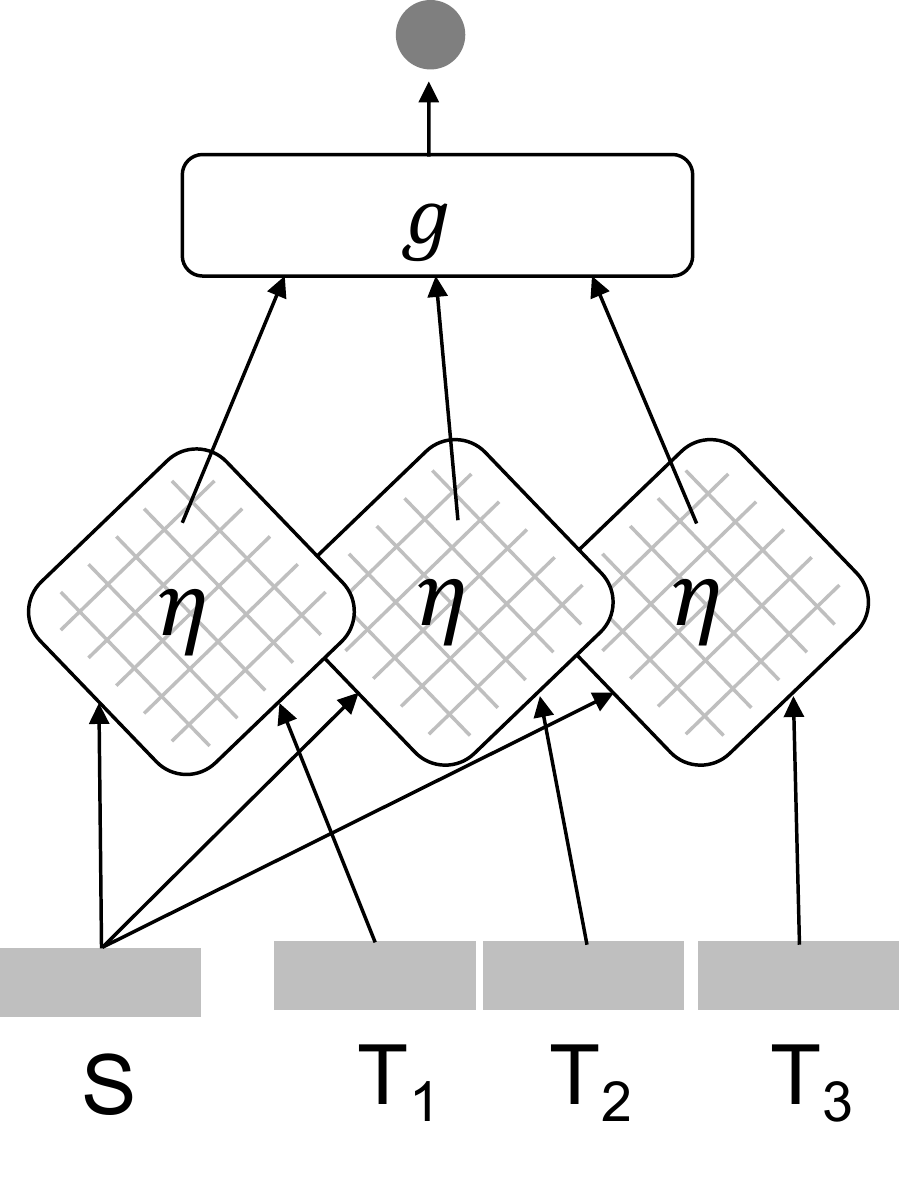}}
        \centerline{\footnotesize{(b) Document Split}}
    \end{minipage}
    \begin{minipage}[t]{0.3\linewidth}
        \centerline{
        \includegraphics[width=0.65\linewidth]{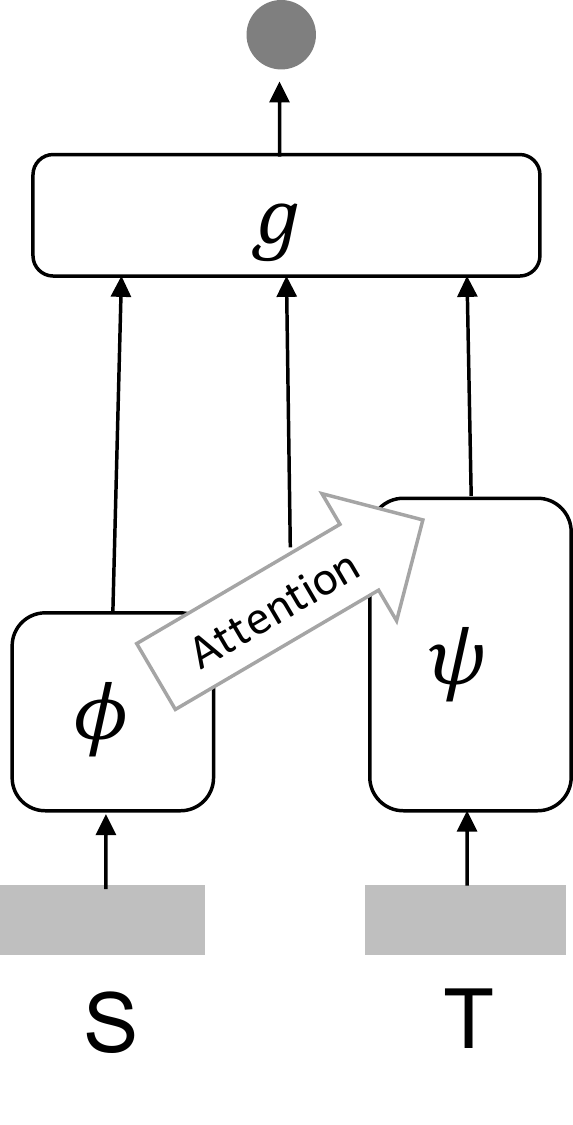}}
        \centerline{\footnotesize{(c) One-way Attention}}
    \end{minipage}
    \caption{Three types of Asymmetric Architecture.}
    \label{fig:asymmetric_arch}
\end{figure}

\begin{itemize}
\item \textit{Query split} is based on the assumption that most queries in ad-hoc retrieval are keyword based, so that we can split the query into terms to match against the document, as illustrated in Figure \ref{fig:asymmetric_arch}(a). A typical model based on this strategy is DRMM~\cite{Guo2016DRMM}. DRMM splits the query into terms and defines the interaction function $\eta$ as the matching histogram mapping between each query term and the document. The evaluation function $g$ consists of two parts, i.e., a feed-forward network for term-level relevance computation and a gating network for score aggregation. Obviously such a process is asymmetric with respect to the query and the document.
K-NRM~\cite{Xiong2017K-NRM} also belongs to this type of approach. It introduces a kernel pooling function to approximate matching histogram mapping to enable end-to-end learning.

\item \textit{Document split} is based on the assumption that a long document could be partially relevant to a query under the scope hypothesis \cite{robertson1976relevance}, so that we split the document to capture fine-grained interaction signals rather than treat it as a whole, as depicted in Figure \ref{fig:asymmetric_arch}(b). A representative model based on this strategy is HiNT~\cite{Fan2018}. In HiNT, the document is first split into passages using a sliding window. The interaction function $\eta$ is defined as the cosine similarity and exact matching between the query and each passage. The evaluation function $g$ includes the local matching layers and global decision layers.  


\item \textit{Joint split}, by its name, uses both assumptions of query split and document split. A typical model based on this strategy is DeepRank~\cite{Pang2017}. Specifically, DeepRank splits the document into  term-centric contexts with respect to each query term. It then defines the interaction function $\eta$ between the query and term-centric contexts in several ways. The evaluation function $g$ includes three parts, i.e., term-level computation, term-level aggregation, and global aggregation.
Similarly, PACRR~\cite{Hui2017} takes the query as a set of terms and splits the document using the sliding window as well as the first-k term window.
\end{itemize}

In addition, in neural ranking models applied for QA, there is another popular strategy leading the asymmetric architecture. We name it \textit{one-way attention mechanism} which typically leverages the question representation to obtain the attention over candidate answer words in order the enhance the answer representation, as illustrated in Figure \ref{fig:asymmetric_arch}(c). For example, IARNN~\cite{IARNN} and CompAgg~\cite{Wang2017ICLR}  get the attentive answer representation sequence that weighted by the question sentence representation.

\subsection{Representation-focused vs. Interaction-focused Architectures}
\label{sec:rep_inter}

\begin{figure}
    \begin{minipage}[t]{0.5\linewidth}
        \centerline{
        \includegraphics[width=0.4\linewidth]{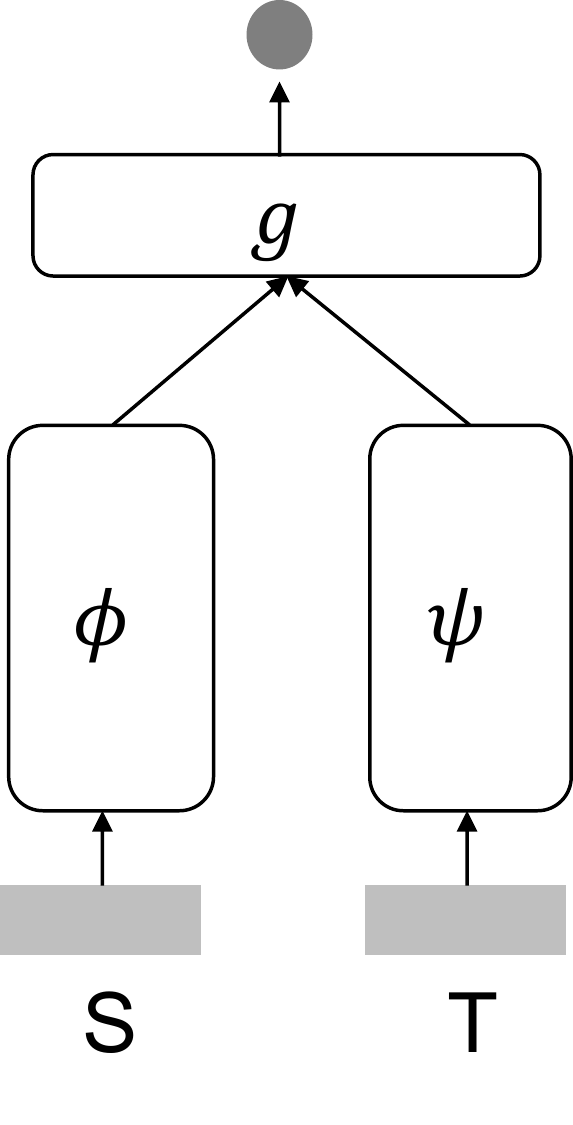}}
        \centerline{\footnotesize{(a) Representation-focused}}
    \end{minipage}
    \begin{minipage}[t]{0.5\linewidth}
        \centerline{
        \includegraphics[width=0.38\linewidth]{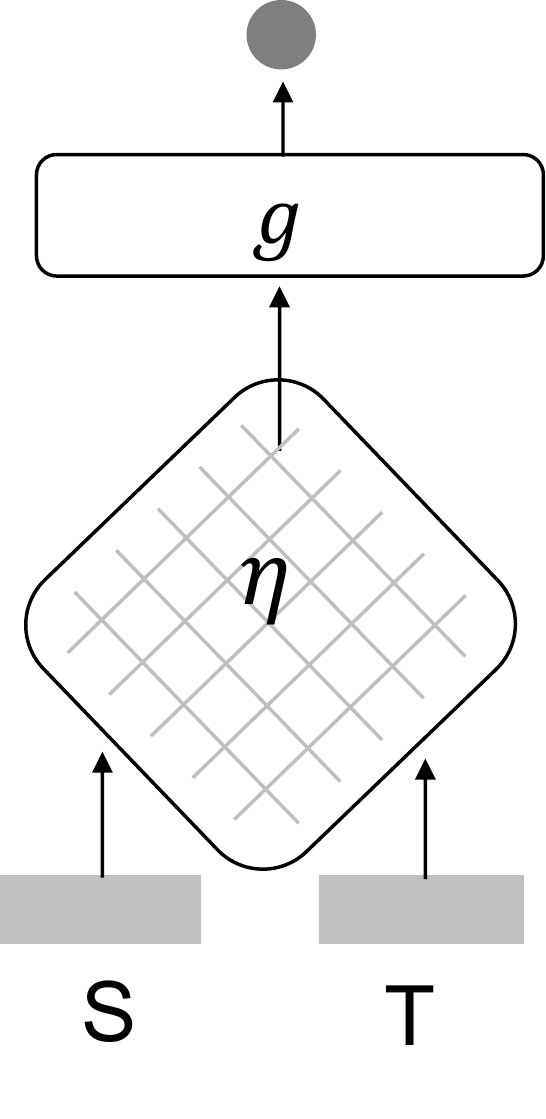}}
        \centerline{\footnotesize{(b) Interaction-focused}}
    \end{minipage}
    \caption{Representation-focused and Interaction-focused Architectures.}
    \label{fig:rep_inter}
\end{figure}

Based on different assumptions over the features (extracted by the representation function $\phi, \psi$ or the interaction function $\eta$) for relevance evaluation, we can divide the existing neural ranking models into another two categories of architectures, namely representation-focused architecture and interaction-focused architecture, as illustrated in Figure \ref{fig:rep_inter}. Besides these two basic categories, some neural ranking models adopt a hybrid way to enjoy the merits of both architectures in learning relevance features.



\textbf{Representation-focused Architecture}: The underlying assumption of this type of architecture is that relevance depends on compositional meaning of the input texts. Therefore, models 
in this category usually define complex representation functions $\phi$ and $\psi$ (i.e., deep neural networks), but no interaction function $\eta$, to obtain high-level representations of the inputs $s$ and $t$, and uses some simple evaluation function $g$ (e.g. cosine function or MLP) to produce the final relevance score. Different deep network structures have been applied for $\phi$ and $\psi$, including fully-connected networks, convolutional networks and recurrent networks.

\begin{itemize}
\item To our best knowledge, DSSM \cite{Huang2013} is the only one that uses the fully-connected network for the functions $\phi$ and $\psi$, which has been described in Section~\ref{sec:symm_nonsymm}.

\item Convolutional networks have been used for $\phi$ and $\psi$ in Arc-I~\cite{Hu2014}, CNTN~\cite{Qiu2015} and  CLSM~\cite{Shen2014}. Take Arc-I as an example, stacked 1D convolutional layers and max pooling layers are applied on the input texts $s$ and $t$ to produce their high-level representations respectively. Arc-I then concatenates the two representations and applies an MLP as the evaluation function $g$. The main difference between CNTN and Arc-I is the function $g$, where the neural tensor layer is used instead of the MLP. The description on CLSM could be found in Section~\ref{sec:symm_nonsymm}.

\item Recurrent networks have been used for $\phi$ and $\psi$ in LSTM-RNN~\cite{Palangi2016} and MV-LSTM~\cite{Wan2016MV-LSTM}. LSTM-RNN uses a one-directional LSTM as $\phi$ and $\psi$ to encode the input texts, which has been described in Section~\ref{sec:symm_nonsymm}. MV-LSTM employs a bi-directional LSTM instead to encode the input texts. Then, the top-k strong matching signals between the two high-level representations are fed to an MLP to generate the relevance score. 
\end{itemize}

By evaluating relevance based on high-level representations of each input text, representation-focused architecture better fits tasks with the global matching requirement \cite{Guo2016DRMM}. This architecture is also more suitable for tasks with short input texts (since it is often difficult to obtain good high-level representations of long texts). Tasks with these characteristics include CQA and AC as shown in Section~\ref{sec:application}. Moreover, models in this category are efficient for online computation, since one can pre-calculate representations of the texts offline once $\phi$ and $\psi$ have been learned.


\textbf{Interaction-focused Architecture}: The underlying assumption of this type of architecture is that relevance is in essence about the relation between the input texts, so it would be more effective to directly learn from interactions rather than from individual representations. Models in this category thus define the interaction function $\eta$ rather than the representation functions $\phi$ and $\psi$, and use some complex evaluation function $g$ (i.e., deep neural networks) to abstract the interaction and produce the relevance score. Different interaction functions have been proposed in literature, which could be divided into two categories, namely non-parametric interaction functions and parametric interaction functions.



\begin{itemize}

\item \textit{Non-parametric interaction functions} are functions that reflect the closeness or distance between inputs without learnable parameters. In this category, some are defined over each pair of input word vectors, such as binary indicator function~\cite{Pang2016, Pang2017}, cosine similarity function~\cite{Pang2016, Yang2016aNMM, Pang2017}, dot-product function~\cite{Pang2016, Pang2017, Fan2018} and radial-basis function \cite{Pang2016}. 
The others are defined between a word vector and a set of word vectors, e.g. the matching histogram mapping in DRMM~\cite{Guo2016DRMM} and the kernel pooling layer in K-NRM~\cite{Xiong2017K-NRM}.  


\item \textit{Parametric interaction functions} are adopted to learn the similarity/distance function from data. For example, Arc-II \cite{Hu2014} uses 1D convolutional layer for the interaction bwteen two phrases. Match-SRNN~\cite{Wan2016MatchSRNN} introduces the neural tensor layer to model complex interactions between input words. Some BERT-based model \cite{yang2019simple} takes attention as the interaction function to learn the interaction vector (i.e., [CLS] vector) between inputs.
In general, parametric interaction functions are adopted when there is sufficient training data since they bring the model flexibility at the expense of larger model complexity. 


\end{itemize}

By evaluating relevance directly based on interactions, the interaction-focused architecture can fit most IR tasks in general. Moreover, by using detailed interaction signals rather than high-level representations of individual texts, this architecture could better fit tasks that call for specific matching patterns (e.g., exact word matching) and diverse matching requirement \cite{Guo2016DRMM}, e.g., ad-hoc retrieval.
This architecture also better fit tasks with heterogeneous inputs, e.g., ad-hoc retrieval and QA, since it circumvents the difficulty of encoding long texts. Unfortunately, models in this category are not efficient for online computation as previous representation-focused models, since the interaction function $\eta$ cannot be pre-calculated until we see the input pair $(s, t)$. Therefore, a better way for practical usage is to apply these two types of models in a ``telescope'' setting, where representation-focused models could be applied in an early search stage while interaction-focused models could be applied later on.

It is worth noting that parts of the interaction-focused architectures have some connections to those in the computer vision (CV) area. For example, the designs of MatchPyramid~\cite{Pang2016} and PACRR~\cite{Hui2017} are inspired by the neural models for the image recognition task. By viewing the matching matrix as a 2-D image, a CNN network is naturally applied to extract hierarchical matching patterns for relevance estimation. These connections indicate that although neural ranking models are mostly applied over textual data, one may still borrow many useful ideas in neural architecture design from other domains. 





\textbf{Hybrid Architecture}: In order to take advantage of both representation-focused and interaction-focused architectures, a natural way is to adopt a hybrid architecture for feature learning. We find that there are two major hybrid strategies to integrate the two architectures, namely combined strategy and coupled strategy. 

\begin{itemize}
    \item Combined strategy is a loose hybrid strategy, which simply adopts both representation-focused and interaction-focused architectures as  sub-models and combines their outputs for final relevance estimation. A representative model using this strategy is DUET~\cite{Mitra2017DUET}. DUET employs a CLSM-like architecture (i.e., a distributed network) and a MatchPyramid-like architecture (i.e., a local network) as two sub-models, and uses a sum operation to combine the scores from the two networks to produce the final relevance score.
\item Coupled strategy, on the other hand, is a compact hybrid strategy. A typical way is to learn representations with attention across the two inputs. Therefore, the representation functions $\phi$ and $\psi$ and the interaction function $\eta$ are compactly integrated. Representative models using this strategy include IARNN~\cite{IARNN} and CompAgg~\cite{Wang2017ICLR}, which have been discussed in the Section \ref{sec:symm_nonsymm}.  Both models learn the question and answer representations via some one-way attention mechanism.
\end{itemize}

\subsection{Single-granularity vs. Multi-granularity Architecture}

The final relevance score is produced by the evaluation function $g$, which takes the features from $\phi$, $\psi$, and $\eta$ as input for estimation. Based on different assumptions on the estimation process for relevance,  we can divide existing neural ranking models into two categories, namely single-granularity models and multi-granularity models.


\begin{figure}
    \begin{minipage}[t]{0.5\linewidth}
        \centerline{
        \includegraphics[width=0.9\linewidth]{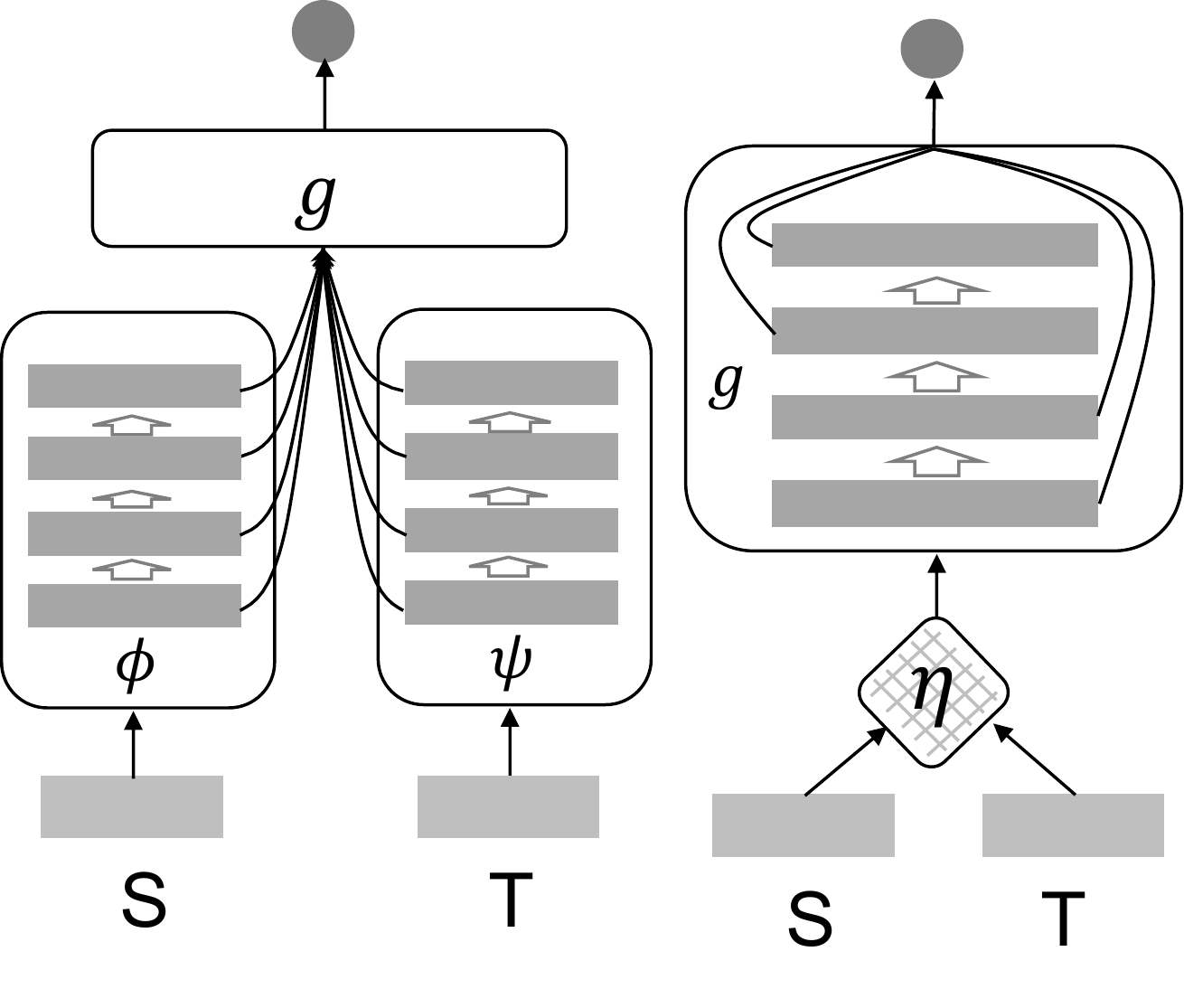}}
        \centerline{\footnotesize{(a) Vertical Multi-granularity}}
    \end{minipage}
    \begin{minipage}[t]{0.5\linewidth}
        \centerline{
        \includegraphics[width=0.9\linewidth]{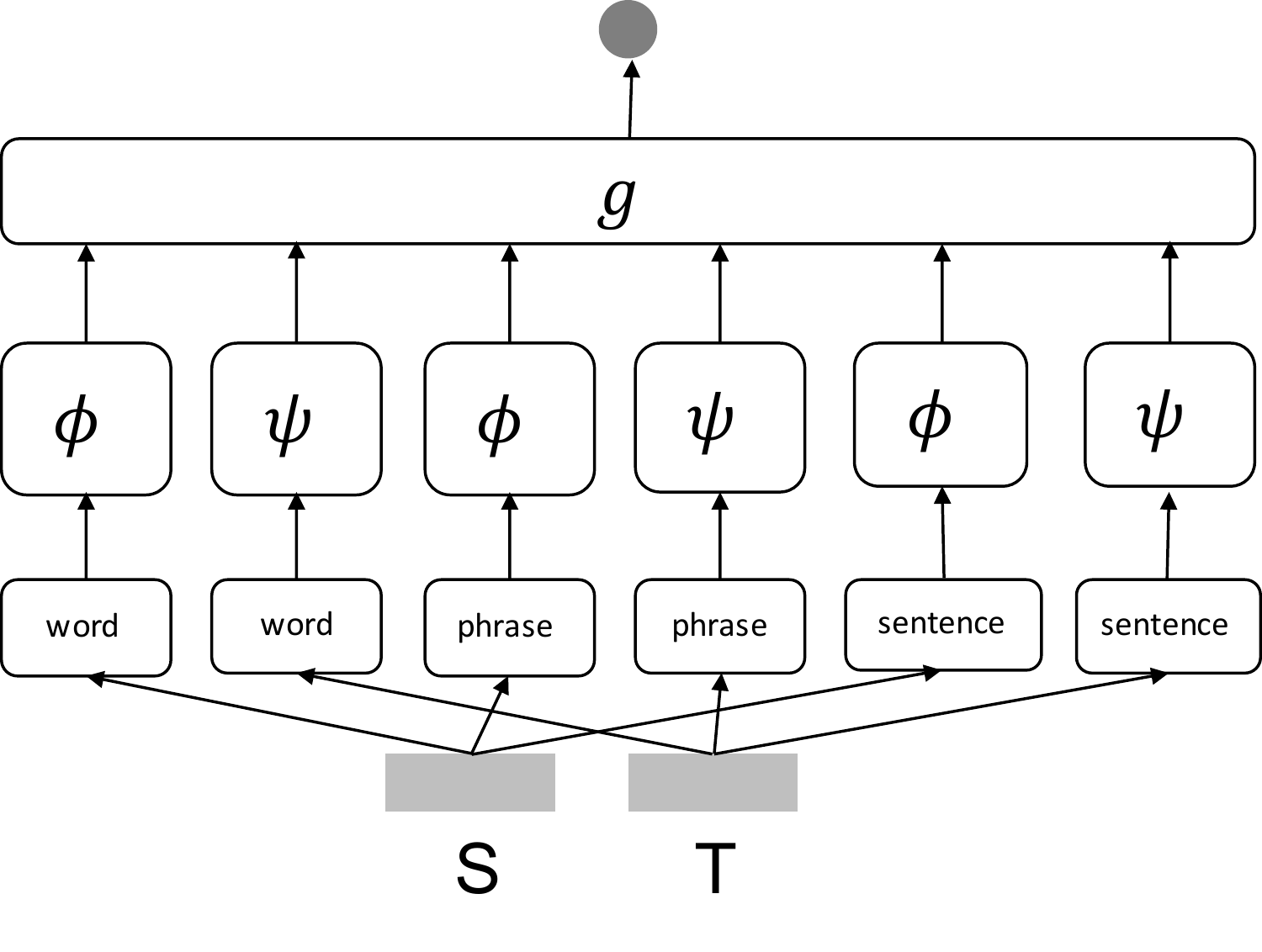}}
        \centerline{\footnotesize{(b) Horizontal Multi-granularity}}
    \end{minipage}
    \caption{Multi-granularity Architectures.}
    \label{fig:multi-granularity}
\end{figure}

\textbf{Single-granularity Architecture}: The underlying assumption of the single-granularity architecture is that relevance can be evaluated based on the high-level features extracted by $\phi$, $\psi$ and $\eta$ from the single-form text inputs. Under this assumption, the representation functions $\phi$, $\psi$ and the interaction function $\eta$ are actually viewed as black-boxes to the evaluation function $g$. Therefore, $g$ only takes their final outputs for relevance computation. Meanwhile, the inputs $s$ and $t$ are simply viewed a set/sequence of words or word embeddings without any additional language structures. 

Obviously, the assumption underlying the single-granularity architecture is very simple and basic. Many neural ranking models fall in this category, with either symmetric (e.g., DSSM and MatchPyramid) or asymmetric (e.g., DRMM and HiNT) architectures, either representation-focused (e.g., ARC-I and MV-LSTM) or interaction-focused (e.g., K-NRM and Match-SRNN).   


\textbf{Multi-granularity Architecture}: The underlying assumption of the multi-granularity architecture is that relevance estimation requires multiple granularities of features, either from different-level feature abstraction or based on different types of language units of the inputs. Under this assumption, the representation functions $\phi$, $\psi$ and the interaction function $\eta$ are no longer black-boxes to $g$, and we consider the language structures in $s$ and $t$. We can identify two basic types of multi-granularity, namely vertical multi-granularity and horizontal multi-granularity, as illustrated in Figure \ref{fig:multi-granularity}.

\begin{itemize}
    \item \textit{Vertical multi-granularity} takes advantage of the hierarchical nature of deep networks so that the evaluation function $g$ could leverage different-level abstraction of features for relevance estimation. For example, In MultigranCNN \cite{Yin2015MultiGranCNN}, the representation functions $\psi$ and $\phi$ are defined as two CNN networks to encode the input texts respectively, and the evaluation function $g$ takes the output of each layer for relevance estimation. MACM \cite{Nie2018multilevel} builds a CNN over the interaction matrix from $\eta$, uses MLP to generate a layer-wise score for each abstraction level of the CNN, and aggregates all the layers' scores for the final relevance estimation. Similar ideas can also be found in MP-HCNN \cite{rao2019multi-perspective} and MultiMatch \cite{nie2018empirical}.
    \item \textit{Horizontal multi-granularity} is based on the assumption that language has intrinsic structures (e.g., phrases or sentences), and we shall consider different types of language units, rather than simple words, as inputs for better relevance estimation. Models in this category typically enhance the inputs by extending it from words to phrases/n-grams or sentences, apply certain single-granularity architectures over each input form, and aggregate all the granularity for final relevance output. For example, in \cite{Huang2017multigran}, a CNN and an LSTM are applied to obtain the character-level, word-level, and sentence-level representations of the inputs, and each level representations are then interacted and aggregated by the evaluation function $g$ to produce the final relevance score. Similar ideas can be found in Conv-KNRM \cite{Dai2018Conv-KNRM} and MIX \cite{Chen2018MIX}.
\end{itemize}

As we can see, the multi-granularity architecture is a natural extension of the single-granularity architecture, which takes into account the inherent language structures and network structures for enhanced relevance estimation. With multi-granularity features extracted, models in this category are expected to better fit tasks that require fine-grained matching signals for relevance computation, e.g., ad-hoc retrieval \cite{Dai2018Conv-KNRM} and QA \cite{Chen2018MIX}. 
However, the enhanced model capability is often reached at the expense of larger model complexity.

\section{Model Learning}

Beyond the architecture, in this section, we review the major learning objectives and training strategies adopted by neural ranking models for comprehensive understadning.

\subsection{Learning objective}



Similar to other LTR algorithms, the learning objective of neural ranking models can be broadly categorized into three groups: \textit{pointwise}, \textit{pairwise}, and \textit{listwise}. 
In this section, we introduce a couple of popular ranking loss functions in each group, and discuss their unique advantages and disadvantages for the applications of neural ranking models in different IR tasks.

\subsubsection{Pointwise Ranking Objective}
The idea of pointwise ranking objectives is to simplify a ranking problem to a set of classification or regression problems. 
Specifically, given a set of query-document pairs $(s_i, t_{i,j})$ and their corresponding relevance annotation $y_{i,j}$, a pointwise learning objective tries to optimize a ranking model by requiring it to directly predict $y_{i,j}$ for $(s_i, t_{i,j})$. 
In other words, the loss functions of pointwise learning objectives are computed based on each $(s,t)$ pair independently. 
This can be formulated as
\begin{equation}
	L(f;\mathcal{S},\mathcal{T},\mathcal{Y}) = \sum_i \sum_j L(y_{i,j},f(s_i, t_{i,j}))
\end{equation}
For example, one of the most popular pointwise loss functions used in neural ranking models is \textit{Cross Entropy}:
\begin{equation}
L(f;\mathcal{S},\mathcal{T},\mathcal{Y}) = -\sum_i \sum_j y_{i,j}\log(f(s_i, t_{i,j})) + (1-y_{i,j})\log(1-f(s_i, t_{i,j}))
\end{equation}
where $y_{i,j}$ is a binary label or annotation with probabilistic meanings (e.g., clickthrough rate), and $f(s_i, t_{i,j})$ needs to be rescaled into the range of 0 to 1 (e.g., with a sigmoid function $\sigma(x) = \frac{1}{1 + \exp(-x)}$).
Example applications include the Convolutional Neural Network for question answering~\cite{Severyn2015}.
There are other pointwise loss functions such as \textit{Mean Squared Error} for numerical labels, but they are more commonly used in recommendation tasks.  

The advantages of pointwise ranking objectives are two-fold. 
First, pointwise ranking objectives are computed based on each query-document pair $(s_i, t_{i,j})$ separately, which makes it simple and easy to scale.
Second, the outputs of neural models learned with pointwise loss functions often have real meanings and value in practice. 
For instance, in sponsored search, a model learned with cross entropy loss and clickthrough rates can directly predict the probability of user clicks on search ads, which is more important than creating a good result list in some application scenarios.

In general, however, pointwise ranking objectives are considered to be less effective in ranking tasks. 
Because pointwise loss functions consider no document preference or order information, they do not guarantee to produce the best ranking list when the model loss reaches the global minimum.
Therefore, better ranking paradigms that directly optimize document ranking based on pairwise loss functions and listwise loss functions have been proposed for LTR problems.

\subsubsection{Pairwise Ranking Objective}

Pairwise ranking objectives focus on optimizing the relative preferences between documents rather than their labels.
In contrast to pointwise methods where the final ranking loss is the sum of loss on each document, pairwise loss functions are computed based on the permutations of all possible document pairs~\cite{chen2009ranking}.
It usually can be formalized as
\begin{equation}
	L(f;\mathcal{S},\mathcal{T},\mathcal{Y}) = \sum_i \sum_{(j,k),y_{i,j}\succ y_{i,k}} L(f(s_i, t_{i,j}) - f(s_i, t_{i,k}))
\end{equation}
where $t_{i,j}$ and $t_{i,k}$ are two documents for query $s_i$ and $t_{i,j}$ is preferable comparing to $t_{i,k}$ (i.e., $y_{i,j} \succ y_{i,k}$).
For instance, a well-known pairwise loss function is \textit{Hinge loss}:
\begin{equation}
L(f;\mathcal{S},\mathcal{T},\mathcal{Y}) = \sum_i \sum_{(j,k),y_{i,j}\succ y_{i,k}}\max(0, 1 - f(s_i, t_{i,j}) + f(s_i, t_{i,k}))
\end{equation}
Hinge loss has been widely used in the training of neural ranking models such as DRMM~\cite{Guo2016DRMM} and K-NRM~\cite{Xiong2017K-NRM}.
Another popular pairwise loss function is the pairwise cross entropy defined as
\begin{equation}
L(f;\mathcal{S},\mathcal{T},\mathcal{Y}) = -\sum_i \sum_{(j,k),y_{i,j}\succ y_{i,k}}\log \sigma (f(s_i, t_{i,j}) - f(s_i, t_{i,k}))
\end{equation}
where $\sigma(x) = \frac{1}{1 + \exp(-x)}$.
Pairwise cross entropy is first proposed in RankNet by Burges et al.~\cite{burges2005learning}, which is considered to be one of the initial studies on applying neural network techniques to ranking problems. 

Ideally, when pairwise ranking loss is minimized, all preference relationships between documents should be satisfied and the model will produce the optimal result list for each query.
This makes pairwise ranking objectives effective in many tasks where performance is evaluated based on the ranking of relevant documents.
In practice, however, optimizing document preferences in pairwise methods does not always lead to the improvement of final ranking metrics due to two reasons:
(1) it is impossible to develop a ranking model that can correctly predict document preferences in all cases;
and (2) in the computation of most existing ranking metrics, not all document pairs are equally important.
This means that the performance of pairwise preference prediction is not equal to the performance of the final retrieval results as a list.
Given this problem, previous studies~\cite{xia2008listwise,taylor2008softrank,burges2010ranknet,ai2018learning} further proposed listwise ranking objectives for learning to rank.
 

\subsubsection{Listwise Ranking Objective}

The idea of listwise ranking objectives is to construct loss functions that directly reflect the model's final performance in ranking.
Instead of comparing two documents each time, listwise loss functions compute ranking loss with each query and their candidate document list together.
Formally, most existing listwise loss functions can be formulated as
\begin{equation}
	L(f;\mathcal{S},\mathcal{T},\mathcal{Y}) = \sum_i L(\{y_{i,j}, f(s_i, t_{i,j})| t_{i,j} \in \mathcal{T}_i\})
\end{equation}
where $\mathcal{T}_i$ is the set of candidate documents for query $s_i$.
Usually, $L$ is defined as a function over the list of documents sorted by $y_{i,j}$, which we refer to as $\pi_i$, and the list of documents sorted by $f(s_i, t_{i,j})$.
For example, Xia et al.~\cite{xia2008listwise} proposed \textit{ListMLE} for listwise ranking as
\begin{equation}
L(f;\mathcal{S},\mathcal{T},\mathcal{Y}) = \sum_{i} \sum_{j=1}^{|\pi_i|}\log P(y_{i,j} | \mathcal{T}_i^{(j)}, f) 
\end{equation} 
where $P(y_{i,j} | \mathcal{T}_i^{(j)}, f)$ is the probability of selecting the $j$th document in the optimal ranked list $\pi_i$ with $f$:
\begin{equation}
P(y_{i,j} | \mathcal{T}_i^{(j)}, f) = \frac{\exp(f(s_i, t_{i,j}))}{\sum_{k=j}^{|\pi_i|} \exp(f(s_i, t_{i,k}))}
\end{equation}
Intuitively, ListMLE is the log likelihood of the optimal ranked list given the current ranking function $f$, but computing log likelihood on all the result positions is computationally prohibitive in practice.
Thus, many alternative functions have been proposed for listwise ranking objectives in the past ten years.
One example is the \textit{Attention Rank} function used in the Deep Listwise Context Model proposed by Ai et al.~\cite{ai2018learning}:
\begin{equation}
\begin{split}
L(f;\mathcal{S},\mathcal{T},\mathcal{Y}) &= -\sum_i\sum_{j}P(t_{i,j}|\mathcal{Y}_i, \mathcal{T}_i )\log P(t_{i,j} | f, \mathcal{T}_i) \\
\text{where~}& P(t_{i,j}|\mathcal{Y}_i, \mathcal{T}_i ) = \frac{\exp(y_{i,j})}{\sum_{k=1}^{|\mathcal{T}_i|} \exp(y_{i,k})}, \\ &P(t_{i,j}|fi, \mathcal{T}_i ) = \frac{\exp(f(s_i, t_{i,j}))}{\sum_{k=1}^{|\mathcal{T}_i|} \exp(f(s_i, t_{i,k}))}
\end{split}
\end{equation}
When the labels of documents (i.e., $y_{i,j}$) are binary, we can further simplify the Attention Rank function with a softmax cross entropy function as
\begin{equation}
	L(f;\mathcal{S},\mathcal{T},\mathcal{Y}) = -\sum_i\sum_{j}y_{i,j}\log \frac{\exp(f(s_i, t_{i,j}))}{\sum_{k=1}^{|\mathcal{T}_i|} \exp(f(s_i, t_{i,k}))}
\end{equation}
The softmax-based listwise ranking loss is one of the most popular learning objectives for neural ranking models such as GSF~\cite{ai2018GSF}.
It is particularly useful when we train neural ranking models with user behavior data (e.g., clicks) under the unbiased learning framework~\cite{ai2018unbiased-1}. 
There are other types of listwise loss functions proposed under different ranking frameworks in the literature~\cite{burges2010ranknet, taylor2008softrank}. 
We ignore them in this paper since they are not popular in the studies of neural IR.

While listwise ranking objectives are generally more effective than pairwise ranking objectives, their high computational cost often limits their applications. They are suitable for the re-ranking phase over a small set of candidate documents.
Since many practical search systems now use neural models for document re-ranking, listwise ranking objectives have become increasingly popular in neural ranking frameworks~\cite{Huang2013, Shen2014, Mitra2017DUET, ai2018learning,  ai2018GSF, ai2018unbiased-1}. 


\subsubsection{Multi-task Learning Objective}

In some cases, the optimization of neural ranking models may include the learning of multiple ranking or non-ranking objectives at the same time.
The motivation behind this approach is to use the information from one domain to help the understanding of information from other domains.
For example, Liu et al.~\cite{liu2015representation} proposed to unify the representation learning process for query classification and Web search by training a deep neural network in which the final layer of hidden variables are used to optimize both a classification loss and a ranking loss.
Chapelle et al.~\cite{chapelle2010multi} proposed a multi-boost algorithm to simultaneously learn ranking functions based on search data collected from 15 countries.

In general, the most common methodology used by existing multi-task learning algorithms is to construct shared representations that are universally effective for ranking in multiple tasks or domains.
To do so, previous studies mostly focus on constructing regularizations or restrictions on model optimizations so that the final model is not specifically designed for a single ranking objective~\cite{liu2015representation,chapelle2010multi}.
Inspired by recent advances on generative adversarial networks (GAN)~\cite{goodfellow2014generative}, Cohen et al.~\cite{cohen2018cross} introduced an adversarial learning framework that jointly learns a ranking function with a discriminator which can distinguish data from different domains.
By training the ranking function to produce representations that cannot be discriminated by the discriminator, they teach the ranking system to capture domain-independent patterns that are usable in cross-domain applications.
This is important as it can significantly alleviate the problem of data sparsity in specific tasks and domains.


\subsection{Training Strategies}

Given the data available for training a neural ranking model, an appropriate training strategy should be chosen. In this section, we briefly review a set of effective training strategies for neural ranking models, including supervised, semi-supervised, and weakly supervised learning.

\emph{Supervised learning} refers to the most common learning strategy in which query-document pairs are labeled. The data can be labeled by expert assessors, crowdsourcing, or can be collected from the user interactions with a search engine as implicit feedback. In this training strategy, it is assumed that a sufficient amount of labeled training data is available. Given this training strategy, one can train the model using any of the aforementioned learning objectives, e.g., pointwise and pairwise. However, since neural ranking models are usually data ``hungry'', academic researchers can only learn models with constrained parameter spaces under this training paradigm due to the limited annotated data. This has motivated researchers to study learning from limited data for information retrieval~\cite{Zamani:2018:LND4IR}.

\emph{Weakly supervised learning} refers to a learning strategy in which the query-document labels are automatically generated using an existing retrieval model, such as BM25. The use of pseudo-labels for training ranking models has been proposed by Asadi et al.~\cite{Asadi:2011}. More recently, Dehghani et al.~\cite{Dehghani:2017} proposed to train neural ranking models using weak supervision and observed up to 35\% improvement compared to BM25 which plays the role of weak labeler. This learning strategy does not require labeled training data. In addition to ranking, weak supervision has shown successful results in other information retrieval tasks, including query performance prediction~\cite{Zamani:2018:neuralqpp}, learning relevance-based word embedding~\cite{Zamani:2017:relwe}, and efficient learning to rank~\cite{Cohen:2018}.

\emph{Semi-supervised learning} refers to a learning strategy that leverages a small set of labeled query-document pairs plus a large set of unlabeled data. Semi-supervised learning has been extensively studied in the context of learning to rank. Preference regularization~\cite{Szummer:2011}, feature extraction using KernelPCA~\cite{Duh:2008}, and pseudo-label generation using labeled data~\cite{Zhang:2016} are examples of such approaches.
In the realm of neural models, fine-tuning weak supervision models using a small set of labeled data \cite{Dehghani:2017} and controlling the learning rate in learning from weakly supervised data using a small set of labeled data~\cite{DBLP:journals/corr/abs-1711-00313} are another example of semi-supervised approaches to ranking. Recently, Li et al.~\cite{Li:2018} proposed a neural model with a joint supervised and unsupervised loss functions. The supervised loss accounts for the error in query-document matching, while the unsupervised loss computes the document reconstruction error (i.e., auto-encoders).


\section{Model Comparison}


In this section, we compare the empirical evaluation results of the previously reviewed neural ranking models on several popular benchmark data sets. We mainly survey and analyze the published results of neural ranking models for the ad-hoc retrieval and QA tasks. Note that sometimes it is difficult to compare published results across different papers - small changes such as  different tokenization, stemming, etc. can lead to significant differences. Therefore, we attempt to collect results from papers that contain comparisons across some of these models performed at a single site for fairness .

\subsection{Empirical Comparison on Ad-hoc Retrieval}
To better understand the performances of different neural ranking models on ad-hoc retrieval, we show the published experimental results on benchmark datasets. Here, we choose three representative datasets for ad-hoc retrieval: 
(1) Robust04 dataset is a standard ad-hoc retrieval dataset where the queries are from TREC Robust Track 2004. 
(2) Gov2$_{\mathit{MQ2007}}$ is an Web Track ad-hoc retrieval dataset where the collection is the Gov2 corpus. The queries are from the Million Query Track of TREC 2007. 
(3) Sougou-Log dataset \cite{Xiong2017K-NRM} is built on query logs sampled from search logs of Sougou.com.  
(4) WT09-14 is the 2009-2014 TREC Web Track, which are based on the ClueWeb09 and ClueWeb12 datasets.
The detailed data statistics can be found in related literature~\cite{Guo2016DRMM, Pang2017, Fan2018, Xiong2017K-NRM,hui2018co}. 

For meaningful comparison, we have tried our best to restrict the reported results to be under the same experimental settings. Specifically, experiments on Robust04 take the title as the query, and all the documents are processed with the Galago Search Engine\footnote{http://www.lemurproject.org/galago.php} \cite{Guo2016DRMM,Zamani:2018:SNRM}. For experiments on the Gov2$_{\mathit{MQ2007}}$ dataset, all the queries and documents are processed using the Galago Search Engine under the same setting as described in \cite{Pang2017, Fan2018}. Besides, the results on the WT09-14 dataset and the Sougou-Log dataset are all from a same paper \cite{hui2018co,Dai2018Conv-KNRM} respectively.


\begin{table}[]
\footnotesize
	\caption{ Overview of previously published results on ad hoc retrieval datasets. The citation in each row denotes the original paper where the method is proposed. The superscripts 1-6 denote that the results are cited from \cite{Guo2016DRMM},\cite{Pang2017},\cite{Fan2018},\cite{hui2018co}, \cite{Zamani:2018:SNRM},  \cite{li2018nprf}, \cite{Dai2018Conv-KNRM} respectively. The subscripts denote the model architecture belongs to (S)ymmetric or (A)symmetric/(R)epresentation-focused or (I)nteraction-focused or (H)ybrid/Singe-(G)ranularity or (M)ulti-granularity. The back slash symbols denote that there are no published results for the specific model on the specific data set in the related literature.}
	\begin{tabular}{l|l|l|l|l|l|l}
		\hline \hline
		\multirow{2}{*}{\diagbox{Model}{Data Set}}                          & \multicolumn{2}{l|}{Robust04}        & \multicolumn{2}{l|}{GOV2$_{\mathit{MQ2007}}$}         & \multicolumn{1}{l|}{WT09-14} & Sougo-Log  \\ \cline{2-7}
		  & MAP & P@20   & MAP & P@10 &ERR@20 & NDCG@1 \\ \hline \hline
		BM25\cite{Robertson:1994:SEA:188490.188561} (1994)$^{1,2}$           & 0.255 & 0.370 & 0.450 & 0.366 & \textbackslash{} & 0.142           \\ \hline
		QL\cite{ponte1998language} (1998)$^{1,4}$   & 0.253     & 0.369    &  \textbackslash{}   & \textbackslash{}  & 0.113 & 0.126 \\ \hline
		RM3\cite{lavrenko2001relevance}(2001)$^5$ & 0.287 & 0.377 &\textbackslash{} &\textbackslash{} &\textbackslash{} &\textbackslash{}  \\\hline
		RankSVM\cite{joachims2002optimizing} (2002)$^{2}$      & \textbackslash{} & \textbackslash{} &  0.464 & 0.381 & \textbackslash{}  & 0.146          \\ \hline 
		LambdaMart\cite{burges2010ranknet} (2010)$^{2}$   &\textbackslash{}  &\textbackslash{}  & 0.468 & 0.384 &\textbackslash{} &\textbackslash{}  \\ \hline 
		\hline
		DSSM\cite{Huang2013} (2013)$^{1,2}_{S/R/G}$
			& 0.095 & 0.171 & 0.409 & 0.352 &\textbackslash{} &\textbackslash{}   \\ \hline
		CDSSM\cite{Shen2014} (2014)$^{1,2}_{S/R/G}$ & 0.067 & 0.125 & 0.364 & 0.291 & \textbackslash{} & 0.144 \\ \hline
		ARC-I\cite{Hu2014} (2014)$^{1,2}_{S/R/G}$  & 0.041 & 0.065 & 0.417 & 0.364 & \textbackslash{} & \textbackslash{}     \\ \hline
		ARC-II\cite{Hu2014} (2014)$^{1,2}_{S/I/G}$  & 0.067 & 0.128 & 0.421 & 0.366 &   \textbackslash{} & \textbackslash{}                \\ \hline
		MP\cite{Pang2016} (2016)$^{1,2,4}_{S/I/G}$    & 0.189 & 0.290 & 0.434 & 0.371 & 0.148  & 0.218         \\ \hline
		Match-SRNN\cite{Wan2016MatchSRNN} (2016)$^{2}_{S/H/G}$ & \textbackslash{} & \textbackslash{}& 0.456 & 0.384 & \textbackslash{} &\textbackslash{} \\\hline
		DRMM\cite{Guo2016DRMM} (2016)$^{1,2,4}_{A/I/G}$         & 0.279 & 0.382 & 0.467 & 0.388 &   0.171 & 0.137       \\ \hline
		Duet\cite{Mitra2017DUET} (2017)$^{3,4}_{A/H/G}$ & \textbackslash{} & \textbackslash{} & 0.474  & 0.398 & 0.134 &\textbackslash{} \\\hline
		DeepRank\cite{Pang2017} (2017)$^{2}_{A/I/G}$& \textbackslash{} & \textbackslash{} & 0.497 & 0.412 &\textbackslash{} &\textbackslash{} \\\hline
		K-NRM\cite{Xiong2017K-NRM} (2017)$^{4}_{A/I/G}$
			& \textbackslash{}& \textbackslash{}& \textbackslash{} & \textbackslash{} & 0.154 & 0.264\\\hline
		PACRR\cite{hui2017position} (2017)$^{6,4}_{A/I/M}$ & 0.254 & 0.363 & \textbackslash{} & \textbackslash{} & 0.191 & \textbackslash{} \\\hline
		Co-PACRR\cite{hui2018co} (2018)$^{4}_{A/I/M}$ & \textbackslash{} & \textbackslash{} & \textbackslash{} & \textbackslash{} & 0.201 & \textbackslash{}\\\hline
		SNRM\cite{Zamani:2018:SNRM} (2018)$^{5}_{S/R/G}$
			& 0.286 & 0.377 & \textbackslash{} & \textbackslash{} &\textbackslash{} & \textbackslash{} \\\hline
	    SNRM+PRF\cite{Zamani:2018:SNRM} (2018)$^{5}_{S/R/G}$
			& 0.297 & 0.395 & \textbackslash{} & \textbackslash{} &\textbackslash{} & \textbackslash{}\\\hline
		CONV-KNRM\cite{Dai2018Conv-KNRM} (2018)$^{4}_{A/I/M}$ & \textbackslash{}& \textbackslash{}& \textbackslash{} & \textbackslash{} & \textbackslash{} & 0.336 \\\hline
		NPRF-KNRM\cite{li2018nprf} (2018)$^{6}_{A/I/G}$ & 0.285 & 0.393 & \textbackslash{} & \textbackslash{} &  \textbackslash{} & \textbackslash{} \\\hline
		NPRF-DRMM\cite{li2018nprf} (2018)$^{6}_{A/I/G}$ & 0.290 & 0.406 & \textbackslash{} & \textbackslash{} &  \textbackslash{}  & \textbackslash{} \\\hline
		HiNT\cite{Fan2018} (2018)$^{3}_{A/I/G}$ & \textbackslash{} & \textbackslash{} & 0.502 & 0.418 & \textbackslash{} & \textbackslash{} \\\hline\hline
	\end{tabular}
	\label{tab:adhoc_exp_results_survey}
\end{table}

Table \ref{tab:adhoc_exp_results_survey} shows an overview of previous published results on ad-hoc retrieval datasets. We have included some well-known probabilistic retrieval models, pseudo-relevance feedback (PRF) models and LTR models as baselines. Based on the results, we have the following observations:
\begin{enumerate}
    \item The probabilistic models (i.e., QL and BM25), although simple, can already achieve reasonably good performance. The traditional PRF model (i.e., RM3) and LTR models (i.e., RankSVM and LambdaMart) with human designed features are strong baselines whose performance is hard to beat for most neural ranking models based on raw texts. However, the PRF technique can also be leveraged to enhance neural ranking models (e.g., SNRM+PRF \cite{Zamani:2018:SNRM} and NPRF+DRMM \cite{li2018nprf} in Table \ref{tab:adhoc_exp_results_survey}), while human designed LTR features can be integrated into neural ranking models \cite{Pang2017,fan2017learning} to improve the ranking performance.  
    \item There seems to be a paradigm shift of the neural ranking model architectures from symmetric to asymmetric and from representation-focused to interaction-focused over time. This is consistent with our previous analysis where asymmetric and interaction-focused structures may fit better with the ad-hoc retrieval task which shows heterogeneity inherently. 
   \item With bigger data size in terms of distinct number of queries and labels (i.e., Sogou-Log $\succ$ GOV2$_{MQ2007}$ $\succ$ WT09-14 $\succ$ Robust04), neural models are more likely to achieve larger performance improvement against non-neural models. As we can see, the best neural models based on raw texts  can significantly outperform LTR models with human designed features on Sogou-Log dataset.
    \item Based on the reported results, in general, we observe that the asymmetric, interaction-focused, multi-granularity architecture can work better than the symmetric, representation-focused, single-granularity architecture on the ad-hoc retrieval tasks. There is one exception, i.e., SNRM on Robust04. However, this model was trained with a large amount of data using the weak supervision strategy, and may not be appropriate to directly compare with those models trained on Robust04 alone.  
\end{enumerate}


\subsection{Empirical Comparison on QA}

In order to understand the performance of different neural ranking models reviewed in this paper for the QA task, we survey the previously published results on three QA data sets, including TREC QA \cite{D07-1003}, WikiQA \cite{wikiqa2015} and Yahoo! Answers \cite{Wan2016MV-LSTM}. TREC QA and WikiQA are answer sentence selection/retrieval data sets and they mainly contain factoid questions, while Yahoo! Answers is an answer passage retrieval data set sampled from the CQA website Yahoo! Answers. The detailed data statistics can be found in related literature \cite{Yu2014DLQA,wikiqa2015,Wan2016MV-LSTM}. 

We have tried our best to report results under the same experimental settings for fair comparison between different methods. Specifically, the results on TREC QA are over the raw version of the data \cite{Rao:2016:NEA:2983323.2983872}\footnote{\url{https://aclweb.org/aclwiki/Question_Answering_(State_of_the_art)}}. WikiQA only has a single version with the same train/ valid/ test data partitions \cite{wikiqa2015}. Yahoo Answers data is the processed version from the same related work \cite{Wan2016MV-LSTM}. Therefore, questions and answer candidates in all the train/valid/test sets used in different surveyed papers are the same, and the results are comparable with each other. 


\begin{table}[]
	\caption{ Overview of previously published results on QA benchmark data sets. The citation in each row denotes the original paper where the method is proposed. The superscripts 1-10 denote that the results are cited from \cite{wikiqa2015}, \cite{Wan2016MatchSRNN}, \cite{Yang2016aNMM}, \cite{IARNN}, \cite{Wan2016MV-LSTM}, \cite{Tay2017SIGIR}, \cite{Wang2017ICLR}, \cite{Tay2018WSDM}, \cite{Yu2014DLQA}, \cite{Chen2018MIX} respectively. The subscripts denote the model architecture belongs to (S)ymmetric or (A)symmetric/(R)epresentation-focused or (I)nteraction-focused or (H)ybrid/Single-(G)ranularity or (M)ulti-granularity. The back slash symbols denote that there are no published results for the specific model on the specific data set in the related literature.}\label{tab:qa_exp_results_survey}
	\footnotesize
	\begin{tabular}{l|l|l|l|l|l|l}
	\hline \hline
		Data Set                          & \multicolumn{2}{c|}{TREC QA}        & \multicolumn{2}{c|}{WikiQA}         & \multicolumn{2}{c}{Yahoo! Answers}   \\ \hline
		Model                             & MAP              & MRR              & MAP              & MRR              & P@1              & MRR              \\ \hline \hline
		BM25\cite{Robertson:1994:SEA:188490.188561} (1994)$^2$           & \textbackslash{} & \textbackslash{} & \textbackslash{} & \textbackslash{} & 0.579            & 0.726            \\ \hline
		LCLR\cite{DBLP:conf/acl/YihCMP13} (2013)$^{1,9}$          & 0.709            & 0.770            & 0.599            & 0.609            & \textbackslash{} & \textbackslash{} \\ \hline
		Word Cnt\cite{Yu2014DLQA} (2014)$^{1,9}$       &  0.571 & 0.627          & 0.489            & 0.492            & \textbackslash{} & \textbackslash{} \\ \hline
		Wgt Word Cnt\cite{Yu2014DLQA} (2014)$^{1,9}$   &  0.596 & 0.652            & 0.510            & 0.513            & \textbackslash{} & \textbackslash{} \\ \hline \hline
		DeepMatch\cite{Lu2013} (2013)$^5_{S/I/G}$       & \textbackslash{} & \textbackslash{} & \textbackslash{} & \textbackslash{} & 0.452            & 0.679            \\ \hline
		CNN\cite{Yu2014DLQA} (2014)$^{1,9}_{S/R/G}$           & 0.569 & 0.661            & 0.619            & 0.628            & \textbackslash{} & \textbackslash{} \\ \hline
		CNN-Cnt\cite{Yu2014DLQA} (2014)$^{1,9}_{S/R/G}$        & 0.711 & 0.785           & 0.652            & 0.665            & \textbackslash{} & \textbackslash{} \\ \hline
		ARC-I\cite{Hu2014} (2014)$^2_{S/R/G}$           & \textbackslash{} & \textbackslash{} & \textbackslash{} & \textbackslash{} & 0.581            & 0.756            \\ \hline
		ARC-II\cite{Hu2014} (2014)$^2_{S/I/G}$          & \textbackslash{} & \textbackslash{} & \textbackslash{} & \textbackslash{} & 0.591            & 0.765           \\ \hline
		CDNN\cite{Severyn2015} (2015)$^3_{S/R/G}$ & 0.746            & 0.808            & \textbackslash{} & \textbackslash{} & \textbackslash{} & \textbackslash{} \\ \hline
		BLSTM\cite{Wang2015BLSTM} (2015)$^3_{S/R/G}$      & 0.713            & 0.791            & \textbackslash{} & \textbackslash{} & \textbackslash{} & \textbackslash{} \\ \hline
		CNTN\cite{Qiu2015} (2015)$^{2,6}_{S/R/G}$            &  0.728 & 0.783 & \textbackslash{} & \textbackslash{} & 0.626            & 0.781            \\ \hline
		MultiGranCNN\cite{Yin2015MultiGranCNN} (2015)$^2_{S/I/M}$    & \textbackslash{} & \textbackslash{} & \textbackslash{} & \textbackslash{} & 0.725            & 0.840             \\ \hline
		LSTM-RNN\cite{Palangi2016} (2016)$^2_{S/R/G}$        & \textbackslash{} & \textbackslash{} & \textbackslash{} & \textbackslash{} & 0.690            & 0.822            \\ \hline
		MV-LSTM\cite{Wan2016MV-LSTM} (2016)$^{2,6}_{S/R/G}$         & 0.708 & 0.782 & \textbackslash{} & \textbackslash{} & 0.766            & 0.869           \\ \hline
		MatchPyramid\cite{Pang2016} (2016)$^2_{S/I/G}$    & \textbackslash{} & \textbackslash{} & \textbackslash{} & \textbackslash{} & 0.764            & 0.867             \\ \hline
		aNMM\cite{Yang2016aNMM} (2016)$^3_{A/I/G}$          & 0.750            & 0.811            & \textbackslash{} & \textbackslash{} & \textbackslash{} & \textbackslash{} \\ \hline
		Match-SRNN\cite{Wan2016MatchSRNN} (2016)$^2_{S/I/G}$   & \textbackslash{} & \textbackslash{} & \textbackslash{} & \textbackslash{} & 0.790            & 0.882            \\ \hline
		IARNN\cite{IARNN} (2016)$^4_{A/H/G}$          & \textbackslash{}           & \textbackslash{}           & 0.734          & 0.742           & \textbackslash{} & \textbackslash{} \\ \hline 
		HD-LSTM\cite{Tay2017SIGIR} (2017)$^6_{S/R/G}$     & 0.750 & 0.815           & \textbackslash{}          & \textbackslash{}         & \textbackslash{} & \textbackslash{} \\ \hline 
		CompAgg\cite{Wang2017ICLR} (2017)$^7_{A/I/G}$     & \textbackslash{} & \textbackslash{} & 0.743           & 0.755          & \textbackslash{} & \textbackslash{} \\ \hline 
		HyperQA\cite{Tay2018WSDM} (2018)$^8_{S/R/G}$     &  0.770 & 0.825           & 0.712 & 0.727        & \textbackslash{} & \textbackslash{} \\ \hline
		MIX\cite{Chen2018MIX} (2018)$^{10}_{S/I/M}$     &  \textbackslash{} & \textbackslash{}          & 0.713 & \textbackslash{}       & \textbackslash{} & \textbackslash{} \\ \hline \hline
	\end{tabular}
\end{table}

Table \ref{tab:qa_exp_results_survey} shows the overview of the published results on the QA benchmark data sets. We include several traditional non-neural methods as baselines. We summarize our observations as follows: 
\begin{enumerate}
    \item Unlike ad-hoc retrieval, symmetric architectures have been more widely adopted in the QA tasks possibly due to the increased homogeneity between the question and the answer, especially for answer sentence retrieval data sets like TREC QA and WikiQA.
    \item Representation-focused architectures have been more adopted on short answer sentence retrieval data sets, i.e., TREC QA and WikiQA, while interaction-focused architectures have been more adopted on longer answer passage retrieval data sets, e.g., Yahoo! Answer. However, unlike ad-hoc retrieval, there seems to be no clear winner between the representation-focused architecture and the interaction-focused architecture on QA tasks.
    \item Similar to ad-hoc retrieval, neural models are more likely to achieve larger performance improvement against non-neural models on bigger data sets. For example, on small data set like TREC QA, feature engineering based methods such as LCLR can achieve very strong performance. However, on large data set like WikiQA and Yahoo! Answers, we can see a clear gap between neural models and non-neural models.
    \item The performance in general increases over time, which might be due to the increased model capacity as well as the adoption of some advanced approaches, e.g., the attention mechanism. For example, IARNN utilizes attention-based RNN models with GRU to get an attentive sentence representation. MIX extracts grammar information and integrates attention matrices in the attention channels to encapsulate rich structural patterns. aNMM adopts attention mechanism to encode question term importance for aggregating interaction matching features.
\end{enumerate}

\section{Trending Topics}
In this section, we discuss several trending topics related to neural ranking models. Some of these topics are important but have not been well addressed in this field, while some are very promising directions for future research. 


\subsection{Indexing: from Re-ranking to Ranking}
Modern search engines take advantage of a multi-stage cascaded architecture in order to efficiently provide accurate result lists to users. In more detail, there can be a stack of rankers, starting from an efficient high-recall model. Learning to rank models are often employed to model the last stage ranker whose goal is to re-rank a small set of documents retrieved by the early stage rankers. The main objective of these learning to rank models is to provide high-precision results. 

Such a multi-stage cascaded architecture suffers from an error propagation problem. In other words, the errors initiated by the early stage rankers are propagated to the last stage. This clearly shows that multi-stage systems are not optimal. However, for efficiency reasons, learning to rank models cannot be used as the sole ranker to retrieve from large collections, which is a disadvantage for such models.

To address this issue, Zamani et al.~\cite{Zamani:2018:SNRM} recently argued that the sparse nature of natural languages enables efficient term-matching retrieval models to take advantage of an \emph{inverted index} data structure for efficient retrieval. Therefore, they proposed a standalone neural ranking model (SNRM) that learns high-dimensional sparse representations for queries and documents. In more detail, this type of model should optimize two objectives: (\textit{i}) a relevance objective that maximizes the effectiveness of the model in terms of the retrieval performance, and (\textit{ii}) a sparsity objective that is equivalent to minimizing $L_0$ of the query and document representations. 
SNRM has shown superior performance compared to competitive baselines and has performed as efficiently as term-matching models, such as TF-IDF and BM25. 

Learning inverted indexes has been also started to be explored in the database community. Kraska et al.~\cite{Kraska:2018} recently proposed to look at indexes as models. For example, a B-Tree-Index can be seen as a function that maps each key to a position of record in a sorted list. They proposed to replace traditional indexes used in databases with the indexes learned using deep learning technologies. Their models demonstrate a significant conflict reduction and memory footprint improvement.

Graph-based hashing and indexing algorithms have also attracted a considerable attention, which could be leveraged to index neural representations for the initial retrieval. For instance, 
Boytsov et al.~\cite{Boytsov:2016} proposed to replace term-matching retrieval models with approximate nearest neighbor algorithms. Van Gysel et al.~\cite{Gysel:2018} used a similar idea to design an unsupervised neural retrieval model, however, their model architecture is not scalable to large document collections.

Moving from re-ranking a small set of documents to retrieving documents from a large collection is a recent research direction with a number of unanswered questions that require further investigation. For example, understanding and interpreting the learned neural representations has yet to be addressed. Furthermore, there is a known trade-off between efficiency and effectiveness in information retrieval systems, however, understanding this trade-off in learning inverted indexes requires further research. In addition, although index compression is a common technique in the search engine industry to reduce the size of the posting lists and improve efficiency, compression of the learned latent indexes is an unexplored area of research. 

In summary, learning to index and developing effective and at the same time efficient retrieval models is a promising direction in neural IR research, however, we still face several open questions in this area.


\subsection{Learning with External Knowledge}

Most existing neural ranking models focus on learning the matching patterns between the two input texts.
In recent years, some researchers have gone beyond matching textual objects by leveraging external knowledge to enhance the ranking performance. These research works can be grouped into two categories: 1) learning with external structured knowledge such as knowledge bases \cite{Liu2018,Xiong2017Word,DBLP:journals/corr/NguyenTSB16,Shen:2018:KAN:3209978.3210081,Song:2018:NCM:3209978.3209996,DBLP:journals/corr/XuLWSW16}; 2) learning with external unstructured knowledge such as retrieved top results, topics or tags \cite{Yang2018,Ghazvininejad2018,Wu2018}. We now briefly review this work.

The first category of research explored improving neural ranking models with semantic information from knowledge bases. Liu et al. \cite{Liu2018} proposed EDRM that incorporates entities in interaction-focused neural ranking models. EDRM first learns the distributed representations of entities using their semantics from knowledge bases in descriptions and types. Then the model matches documents to queries with both bag-of-words and bag-of-entities. Similar approaches were proposed by Xiong et al. \cite{Xiong2017Word}, which also models queries and documents with word-based representations and entity-based representations. Nguyen et al. \cite{DBLP:journals/corr/NguyenTSB16} proposed combining distributional semantics learned through neural networks and symbolic semantics held by extracted concepts or entities from text knowledge bases to enhance the learning algorithm of latent representations of queries and documents. Shen et al. \cite{Shen:2018:KAN:3209978.3210081} proposed the KABLSTM model, which leverages external knowledge from knowledge graphs to enrich the representational learning of QA sentences. Xu et al. \cite{DBLP:journals/corr/XuLWSW16} designed a Recall gate, where domain knowledge can be transformed into the extra global memory of LSTM, with the aim of enhancing LSTM by cooperating with its local memory to capture the implicit semantic relevance between sentences within conversations.

Beyond structured knowledge in knowledge bases, other research has explored how to integrate external knowledge from unstructured texts, which are more common for information on the Web. Yang et al. \cite{Yang2018}  studied response ranking in information-seeking conversations and proposed two effective methods to incorporate external knowledge into neural ranking models with pseudo-relevance feedback (PRF) and QA correspondence knowledge distillation. They proposed to extract the ``correspondence'' regularities between question and answer terms from retrieved external QA pairs as external knowledge to help response selection. Another representative work on integrating unstructured knowledge into neural ranking models is the KEHNN model proposed by Wu et al. \cite{Wu2018}, which defined prior knowledge as topics, tags, and entities related to the text pair. KEHNN represents global context obtained from external textual collection, and then exploits a knowledge gate to fuse the semantic information carried by the prior knowledge into the representation of words. Finally, it generates a knowledge enhanced representation for each word to construct the interaction matrix between text pairs. 


In summary, learning with external knowledge is an active research area related to neural ranking models. More research efforts are needed to improve the effectiveness of neural ranking models with distilled external knowledge and to understand the role of external knowledge in ranking tasks.

\subsection{Learning with Visualized Technology}

We have discussed many neural ranking models in this survey under the textual IR scenario. 
There have also been a few studies showing that the textual IR problem could be solved visually. The key idea is that we can construct the matching between two inputs as an image so that we can leverage deep neural models to estimate the relevance based on visual features. The advantage of the matching image, compared with traditional matching matrix, is that it can keep the layout information of the original inputs so that many useful features such as spatial proximity, font size and colors could be modeled for relevance estimation. This is especially useful when we consider ad-hoc retrieval tasks on the Web where pages are often well designed documents with rich layout information. 

Specifically, Fan et al.~\cite{fan2017learning} proposed a visual perception model (ViP) to perceive visual features for relevance estimation. They first rendered the Web pages into query-independent snapshots and query-dependent snapshots. Then, the visual features are learned through a combination of CNN and LSTM, inspired by users' reading behaviour. The results have demonstrated the effectiveness of learning the visual features of document for ranking problems. 
Zhang et al.~\cite{zhang2018relevance} proposed a joint relevance estimation model which learns visual patterns, textual semantics and presentation structures jointly from screenshots, titles, snippets and HTML source codes of search results. Their results have demonstrated the viability of the visual features in search result page relevance estimation.
Recently, Akker et al.~\cite{akker2019vitor} built a dataset for the LTR task with visual features, named Visual learning TO Rank (ViTOR). The ViTOR dataset consists of visual snapshots, non-visual features and relevance judgments for ClueWeb12 webpages and TREC Web Track queries. Their results have demonstrated that visual features can significantly improve the LTR performance.

In summary, solving the textual ranking problem through visualized technology is a novel and interesting direction. In some sense, this approach simulates human behavior as we also judge relevance through visual perception. The existing work has only demonstrated the effectiveness of visual features in some relevance assessment tasks. However, more research is needed to understand what can be learned by such visualized technology beyond those text-based methods, and what IR applications could benefit from such models. 



\subsection{Learning with Context}
Search queries are often short and cannot precisely express the underlying information needs. To address this issue, a common strategy is to exploit \emph{query context} to improve the retrieval performance. Different types of query context have been explored in the literature:
\begin{itemize}
    \item Short-term history: the user past interactions with the system in the current search session~\cite{Shen:2005,Ustinovskiy:2013,Xiang:2010}.
    \item Long-term history:  the historical information of the user's queries that is often used for web search personalization~\cite{Bennett:2012,Matthijs:2011}.
    \item Situational context: the properties of the current search request, independent from the query content, such as location and time \cite{Bennett:2011,Zamani:2017:www}.
    \item (Pseudo-) relevance feedback: explicit, implicit, or pseudo relevance signals for a given query can be used as the query context to improve the retrieval performance.
\end{itemize}

Although query context has been widely explored in the literature, incorporating query context into neural ranking models is relatively less studied. Zamani et al.~\cite{Zamani:2017:www} proposed a deep and wide network architecture in which the deep part of the model learns abstract representations for contextual features, while the wide part of the model uses raw contextual features in binary format in order to avoid information loss as a result of high-level abstraction. 
Ahmad et al.~\cite{Ahmad:2018} incorporated short-term history information into a neural ranking model by multi-task training of document ranking and query suggestion. Short- and long-term history have been also used by Chen et al.~\cite{Chen:2018} for query suggestion. 

In addition, learning high-dimensional representation for pseudo-relevance feedback has been also studied in the literature. In this area, embedding-based relevance models~\cite{Zamani:2016:ICTIR} extend the original relevance models~\cite{lavrenko2001relevance} by considering word embedding vectors. The word embedding vectors can be obtained from self-supervised algorithms, such as word2vec~\cite{Mikolov2013}, or weakly supervised algorithms, such as relevance-based word embedding~\cite{Zamani:2017:relwe}. Zamani et al.~\cite{Zamani:2016:CIKM} proposed RFMF, the first pseudo-relevance feedback model that learns latent factors from the top retrieved document. RFMF uses non-negative matrix factorization for learning latent representations for words, queries, and documents. Later on, Li et al.~\cite{li2018nprf} extended existing neural ranking models, e.g., DRMM~\cite{Guo2016DRMM} and KNRM~\cite{Xiong2017K-NRM}, by a neural pseudo-relevance feedback approach, called NPRF. The authors showed that in many cases extending a neural ranking model with NPRF leads to significant improvements. Zamani et al.~\cite{Zamani:2018:SNRM} also made a similar conclusion by extending SNRM with pseudo-relevance feedback.


In summary, with the emergence of interactive or conversational search system, context-aware ranking would be an indispensable technology in these scenarios. These exist several open research questions on how to incorporate query context information in neural ranking models. More research work is expected in this direction in the short future.


\subsection{Neural Ranking Model Understanding}

Deep learning techniques have been widely criticized as a ``black box'' which produces good results but no problem insights and explanations.
Thus, how to understand and explain neural models has been an important topic in both Machine Learning and IR communities.
To the best of our knowledge, the explainability of neural ranking models has not been fully studied. Instead, there have been a few papers on analyzing and understanding the empirical effect of different model components in IR tasks. 

For example, Pang et al.~\cite{pang2016study} conducted an extensive analysis on the MatchPyramid model in ad-hoc retrieval and compared different kernals, pooling sizes, and similarity functions in terms of retrieval performance.
Cohen et al.~\cite{Cohen2018} extracted the internal representations of neural ranking models and evaluated their effectiveness in four natural language processing tasks.
They find that topical relevance information is usually captured in the high-level layers of a neural model.
Nie et al.~\cite{Nie2018} conducted empirical studies on the interaction-based neural ranking model to understand what have been learned in each neural network layer.
They also notice that low-level network layers tend to capture detailed text information while high-level layers tend to have higher topical information abstraction.

While the paradigms of analyzing neural ranking models often rely on a deep understanding of specific model structure, Cohen et al.~\cite{cohen2016adaptability} argue that there are some general patterns of which types of neural models are more suitable for each IR task.
For example, retrieval tasks with fine granularity (e.g., factoid QA) usually need higher levels of information abstraction and semantic matching, while retrieval tasks with coarse granularity (e.g., document retrieval) often rely on the exact matching or interaction between query words and document words.

Overall, the research area on the explainability of neural ranking models is largely unexplored up till now. Some skepticism about neural ranking models is also related to this, e.g., what new things can be learned by neural ranking models? It is a very challenging and promising direction for researchers in neural IR.

\section{Conclusion}
The purpose of this survey is to summarize the current research status on neural ranking models, analyze the existing methodologies, and gain some insights for future development. We introduced a unified formulation over the neural ranking models, and reviewed existing models based on this formulation from different dimensions under model architecture and model learning. For model architecture analysis, we reviewed existing models to understand their underlying assumptions and major design principles, including how to treat the inputs, how to consider the relevance features, and how to make evaluation. For model learning analysis, we reviewed popular learning objectives and training strategies adopted for neural ranking models. To better understand the current status of neural ranking models on major applications, we surveyed published empirical results on the ad-hoc retrieval and QA tasks to conduct a comprehensive comparison. In addition, we discussed several trending topics that are important or might be promising in the future.  

Just as there has been an explosion in the development of many deep learning based methods, research on neural ranking models has increased rapidly and broadened in terms of applications. We hope this survey can help researchers who are interested in this direction, and will motivate new ideas by looking at past successes and failures. Neural ranking models are part of the broader research field of neural IR, which is a joint domain of deep learning and IR technologies with many opportunities for new research and applications. We are expecting that, through the efforts of the community, significant breakthroughs will be achieved in this domain in the near future, similar to those happened in computer vision or NLP.


\section{Acknowledgments}
This work was funded by the National Natural Science Foundation of China (NSFC) under Grants No. 61425016 and 61722211, and the Youth Innovation Promotion Association CAS under Grants No. 20144310. This work was supported in part by the UMass Amherst Center for Intelligent Information Retrieval and in part by NSF IIS-1715095. Any opinions, findings and conclusions or recommendations expressed in this material are those of the authors and do not necessarily reflect those of the sponsor.


\bibliography{bibliography}

\end{document}